\documentclass[superscriptaddress, amsmath, nofootinbib, twocolumn]{revtex4-1}
\usepackage{microtype} 
\usepackage{graphicx}
\usepackage{xcolor}
\usepackage{orcidlink}
\usepackage{enumerate}
\usepackage{enumitem}
\usepackage{ulem}
\definecolor{linkcolor}{rgb}{0.0,0.3,0.5}
\definecolor{linkcolor}{rgb}{0.0,0.3,0.5}
\definecolor{mypurple}{RGB}{143, 116, 210}

\newcommand{\phicen}{|\phi_{\mathrm{c}}|}

\graphicspath{{Figures/}}

\newcommand{\hias}{School of Fundamental Physics and Mathematical Sciences, Hangzhou Institute for Advanced Study, University of Chinese Academy of Sciences, Hangzhou 310024, China}

\newcommand{\kcl}{Theoretical Particle Physics and Cosmology Group,
Physics Department, King's College London, Strand, London WC2R 2LS,
United Kingdom}

\begin{document}

\title{Timing-Window Mechanism for Chain-Like Transients in Collisions of Radially Excited Boson Stars}

% Massive Boson Stars: Stability and Head-on Merger Gravitational Wave Energy Emission

%The Missing Massive Sector: Massive Boson Stars —- Stability and Head-on Merger Gravitational-Wave Energy Emission

\author{Bo-Xuan Ge
\orcidlink{0000-0003-0738-3473}}
%\email{bo-xuan.ge@hotmail.com}
\email{bo-xuan.ge@ucas.ac.cn}
\affiliation{\hias}
\affiliation{\kcl}
%---------------------------------------

\begin{abstract}
We show that chain-like transients in head-on collisions of radially excited boson stars are controlled by the binary collision time, not by radial excitation alone. 
For selected configurations with radial excitation number $n=2$ and
repulsive quartic self-interaction coupling $\lambda=400$, isolated
evolutions define transient breathing windows that serve as reference
clocks. Numerical-relativity simulations
show that visible chains form only when the collision time is compatible
with the isolated breathing clock. A separation scan shifts the collision time relative to the same clock,
providing direct evidence for the timing-window mechanism.
An additional fixed-separation check at \(\lambda=500\) shows the same
event ordering, indicating that the observed pattern is not unique to the
fiducial self-interaction strength.
\end{abstract}

\maketitle

%================================================
\section{Introduction}
\label{sec:intro}
%================================================
\begin{figure}[t]
\centering
\includegraphics[width=\columnwidth]{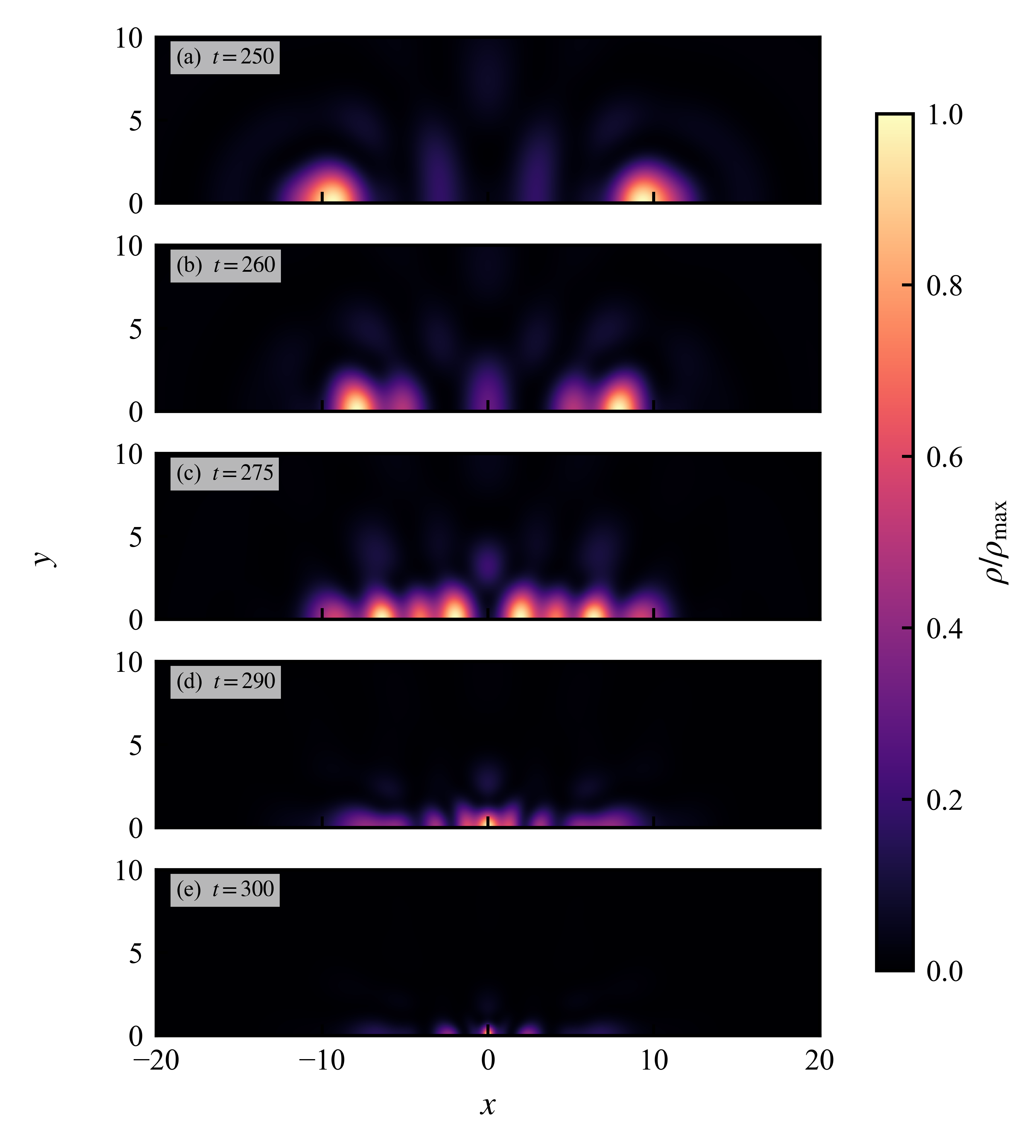}
\caption{
Representative density snapshots of a visible-chain event for the binary
with \(\phicen=0.0875\) and \(D=80\). The panels show the normalized
energy-density morphology in the symmetry-reduced half-plane at
\(t=250,260,275,290\), and \(300\). The sequence illustrates the
formation, visible stage, and subsequent weakening of the chain-like
morphology.
}
\label{fig:chain_snapshot}
\end{figure}
%------------------------------------------------

Boson stars are self-gravitating configurations of complex scalar fields
and provide one of the simplest examples of compact objects without an
ordinary matter surface or an event horizon. Since the original
constructions of mini boson stars~\cite{Kaup:1968zz,Ruffini:1969qy}
and the realization that self-interactions can support astrophysically
massive configurations~\cite{Colpi:1986ye}, they have been studied as
relativistic solitonic objects, compact-object alternatives, and
possible black-hole mimickers. Their isolated structure, stability, and
phenomenology have been explored in a broad range of models
~\cite{Colpi:1986ye, Seidel:1990jh,
Kobayashi:1994qi, Ryan:1996nk, Schunck:1996he, Balakrishna:1997ej,
Yoshida:1997qf, Schunck:1999zu, Schunck:2003kk, Balakrishna:2006ru,
Balakrishna:2007mr, Hartmann:2012da, Siemonsen:2020hcg,
Evstafyeva:2025mvx, Marks:2025jpt, Evstafyeva:2023kfg, Marks:2025xxv,
Marks:2025jit, Ma:2024olw, Ding:2023syj, Liang:2022mjo,
Zhang:2023qxf, Zhang:2025xnl, Ning:2026qxs, deSa:2025nsx, Herdeiro:2025lwf,
Herdeiro:2024myz, Ildefonso:2023qty, Brito:2023fwr}, and their formation
has been studied through scalar-field collapse, gravitational cooling,
and relaxation scenarios~\cite{Seidel:1993zk, Schunck:1999pm,
Sanchis-Gual:2019ljs, Siemonsen:2023hko}.

Fully relativistic simulations of boson-star binaries have revealed an
equally rich nonlinear phenomenology. Depending on compactness,
self-interaction, relative phase, and initial configuration, bosonic
compact-object encounters can produce long-lived remnants, dispersal,
prompt or delayed collapse, and strongly
time-dependent scalar-field structures~\cite{Palenzuela:2006wp,
Palenzuela:2007dm, Palenzuela:2017kcg, Helfer:2021brt,
Sanchis-Gual:2020mzb, Bezares:2022obu, 
Croft:2022bxq, Sanchis-Gual:2022zsr, Evstafyeva:2022bpr,
Siemonsen:2023age, Ge:2024fum, Evstafyeva:2024qvp,
Brito:2025rld, Jaramillo:2022zwg, Damour:2025oys, Ge:2026kkq}. These studies show
that bosonic compact-object dynamics is controlled not only by the
equilibrium sequence, but also by how scalar-field structure is driven
out of equilibrium.

Radially excited boson stars provide a natural setting for probing this
connection. Unlike fundamental configurations, they contain scalar-field
nodes and therefore possess internal length scales that can be rearranged
during evolution. Their stability is also subtle: configurations that
appear long lived in restricted or effectively spherical dynamics may
become unstable once multidimensional degrees of freedom are allowed.
Recent simulations have shown that head-on collisions of excited boson
stars can produce transient chain-like and ring-like
morphologies~\cite{Brito:2025rld}. However, radial excitation alone does
not explain when such structures form, why they are visible only in some
cases, or why they disappear.

In this work we show that visible chain-like transients in collisions
of radially excited boson stars are controlled by the binary collision
time. The relevant clock is the breathing window identified from the
corresponding isolated evolution. This window is not used as a sharp
boundary, but as a reference for the stage during which the internal
shell structure of an excited star is dynamically rearranged. Here we use the term ``breathing'' in a broader operational sense than is
customary in the boson-star and oscillaton literature. We do not mean a
long-lived, periodically pulsating self-gravitating configuration. Rather,
the breathing stage considered here denotes the finite-duration,
perturbation-induced radial oscillation and internal restructuring observed
during the evolution of the excited equilibrium configuration, typically
before the collapse diagnostic is reached. If the
collision occurs well before this stage, the breathing structure has not
yet developed; if it occurs well after this stage, the opportunity for
visible chain formation has been missed. Visible chains therefore appear
only when the binary collision time is compatible with the relevant
breathing window.

We demonstrate this picture for selected configurations along the
radially excited \(n=2\), \(\lambda=400\) sequence of self-interacting
boson stars, using the axisymmetric numerical framework developed and
validated in Refs.~\cite{Ge:2024fum,Ge:2025btw,Ge:2024itl}. Isolated
evolutions first identify the breathing windows of the sampled
configurations. We then collide equal-mass, in-phase copies of the same
configurations at fixed \(D=80\) and \(v=0.1\). Since different sampled
configurations have different intrinsic breathing windows, the same
binary setup can be compatible with the relevant window for some cases
and incompatible with it for others. Visible chains form when the
collision time lies within, or sufficiently close to, the corresponding
breathing window, as illustrated in Fig.~\ref{fig:chain_snapshot}.

As a direct timing test, we fix one representative central amplitude and
vary the initial separation, thereby shifting the collision time while
leaving the individual stellar configuration unchanged. The resulting
turn-on and turn-off of visible chain formation confirms that the chain
is controlled by the relative timing between the binary collision and the
breathing window of the selected excited configuration. An additional
fixed-separation check at \(\lambda=500\), reported in the Supplemental
Material, shows the same event ordering. Resolution checks and a
three-dimensional code comparison further show that the representative
visible-chain morphology is not a resolution artifact or a consequence
of the symmetry-reduced visualization.

%================================================
\section{Setup and observables}
\label{sec:Setup}
%================================================

We consider a complex scalar field \(\phi\) minimally coupled to gravity,
with action
\begin{equation}
S=\int d^4x\sqrt{-g}
\left[
\frac{R}{16\pi}
-\frac{1}{2}g^{\mu\nu}\nabla_{\mu}\bar{\phi}\nabla_{\nu}\phi
-\frac{1}{2}V(|\phi|^2)
\right],
\label{eq:action}
\end{equation}
where an overbar denotes complex conjugation. We use the quartic
self-interaction potential
\begin{equation}
V(|\phi|^2)
=
m^2|\phi|^2+\frac{\lambda}{2}|\phi|^4 ,
\label{eq:quartic_potential}
\end{equation}
and work in units \(G=c=1\), with \(m=1\). In the convention of Eq.~\eqref{eq:quartic_potential}, $\lambda>0$ corresponds to a
repulsive quartic self-interaction. We restrict the present analysis
to this case.

Isolated boson-star solutions are constructed with the harmonic ansatz
\begin{equation}
\phi(t,r)=|\phi(r)|e^{i\omega t},
\label{eq:harmonic_ansatz}
\end{equation}
where \(\omega\) is the scalar-field frequency and
\(\phicen\equiv |\phi(0)|\) denotes the central scalar-field amplitude.
The resulting ordinary differential eigenvalue problem is solved by the
standard shooting construction for massive boson stars, following
Ref.~\cite{Helfer:2021brt}. The radial excitation number \(n\) is defined
by the number of nodes of the scalar profile. 

Both isolated and binary evolutions are performed with our
{\sc GRChombo}-based numerical-relativity code~\cite{Andrade:2021rbd} using the CCZ4 formulation and
adaptive mesh refinement. The configurations considered here are
axisymmetric. We therefore use the modified Cartoon method to evolve the
corresponding three-dimensional axisymmetric dynamics on a
two-dimensional computational domain. The general Cartoon framework
follows Refs.~\cite{Cook:2016soy,Cook:2016qnt}, and its implementation
for boson-star evolutions follows our previous work
~\cite{Ge:2024fum,Ge:2025btw,Ge:2024itl}. Further numerical details are
given in the Supplemental Material~\cite{SupplementalMaterial}.

Binary initial data are constructed from two equal-mass, in-phase excited
boson stars using the improved superposition method of
Refs.~\cite{Helfer:2021brt,Helfer:2018vtq}. The binaries are nonspinning
and undergo head-on collisions with initial separation \(D\) and equal
inward velocity \(v\). We first compare selected configurations evolved
with the same initial separation and velocity, and then perform a direct timing test by fixing \(\phicen\) and
varying \(D\).

The timing analysis combines isolated and binary evolutions. For each
selected isolated configuration, we identify the onset and end of the
large-amplitude breathing stage,
\begin{equation}
t_{\rm br,start}^{\rm iso},
\qquad
t_{\rm br,end}^{\rm iso},
\end{equation}
and define the isolated breathing window by
\begin{equation}
\Delta t_{\rm br}^{\rm iso}
=
t_{\rm br,end}^{\rm iso}
-
t_{\rm br,start}^{\rm iso}.
\label{eq:breathing_window}
\end{equation}
This window is used as a reference clock for the internal restructuring
of the excited star. We also record the isolated diagnostic BH time
\(t_{\rm BH}^{\rm iso}\), defined operationally as the first output time
at which
\begin{equation}
\chi_{\min}(t) < 10^{-3},
\label{eq:BH_diagnostic_threshold}
\end{equation}
where \(\chi_{\min}\) is the minimum value of the conformal factor on the
computational domain. This threshold is used only as a diagnostic
criterion for the timing analysis.

For binary evolutions we extract four characteristic times. The
collision time \(t_{\rm collision}^{\rm binary}\) is identified from the
formation of a common remnant using the bridge diagnostic. The chain
onset time \(t_{\rm chain}^{\rm binary}\) is the first time at which a
chain label is sustained. The binary diagnostic BH time
\(t_{\rm BH\,formed}^{\rm binary}\) is the first time at which the
binary evolution reaches the same conformal-factor diagnostic threshold,
Eq.~\eqref{eq:BH_diagnostic_threshold}. Finally, we record
\(t(S_{\rm chain,valid}^{\max})\), the time at which the chain morphology
score reaches its maximum in the valid pre-BH classification channel.
The detailed bridge criterion, sustainment condition, and
morphology-score construction are given in the Supplemental Material.

As an auxiliary measure of the remnant shape, we compute the
density-weighted second-moment tensor
\begin{equation}
\mathcal{Q}_{ij}(t)
=
\frac{
\int \rho(t,\mathbf{x})
\left[x_i-X_i(t)\right]\left[x_j-X_j(t)\right]
\sqrt{\gamma}\,d^3x
}{
\int \rho(t,\mathbf{x})\sqrt{\gamma}\,d^3x
},
\label{eq:shape_tensor}
\end{equation}
where \(\rho\) is the energy density, \(\gamma\) is the determinant of the
spatial metric, and \(X_i(t)\) is the density-weighted centre of the
configuration. In the axisymmetric head-on channel we use
\begin{equation}
A_{\rm sh}(t)
=
\frac{\mathcal{Q}_{xx}-\mathcal{Q}_{\perp\perp}}
{\mathcal{Q}_{xx}+\mathcal{Q}_{\perp\perp}},
\qquad
\mathcal{Q}_{\perp\perp}
=
\frac{1}{2}
\left(
\mathcal{Q}_{yy}+\mathcal{Q}_{zz}
\right),
\label{eq:shape_anisotropy}
\end{equation}
to track the elongation and redistribution of the remnant. This quantity
is used as an auxiliary morphology diagnostic rather than as the sole
criterion for chain formation.

The visible-chain classification is based on a morphology score
\(S_{\rm chain}\), constructed from resolved axis-supported density
peaks, common-remnant connectivity, and suppression of off-axis weight.
A case is counted as a visible chain only if a sustained chain-like
morphology is present in the valid pre-BH classification channel and
\begin{equation}
S_{\rm chain,valid}^{\max}\geq S_{\rm vis},
\qquad
S_{\rm vis}=2 .
\label{eq:visible_chain_threshold}
\end{equation}
This threshold removes weak or visually ambiguous chain-like candidates.
The detailed definitions of the node filters, bridge diagnostic,
morphology labels, and visible-chain criterion are given in the
Supplemental Material. These thresholds are operational choices selected to suppress
isolated diagnostic fluctuations and to distinguish weak candidates
from clearly resolved events. Representative variations shift the
extracted event times but do not alter the event ordering or the
visible-chain classification.

%================================================
\section{Timing-window evidence from the \texorpdfstring{\(\phi_c\)}{phic} comparison}
\label{sec:amplitude_scan}
%================================================
\begin{figure}[t]
\centering
\includegraphics[width=\columnwidth]{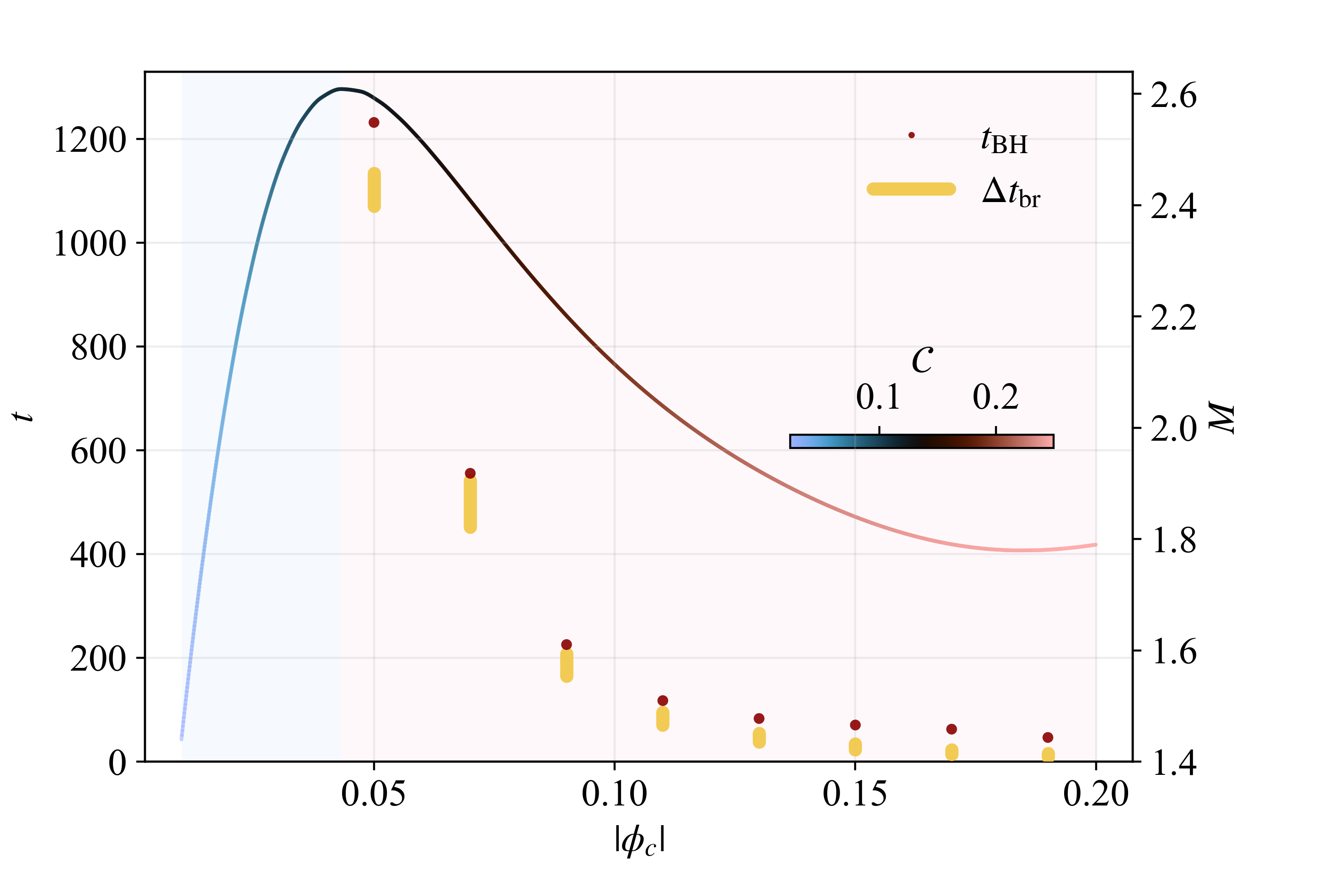}
\caption{
Isolated clocks for the $n=2$, $\lambda=400$ sequence. The curve shows
$M(\phicen)$, with colour indicating compactness. Red points mark
$t_{\rm BH}^{\rm iso}$, defined by $\chi_{\min}<10^{-3}$, and yellow bars
show the breathing interval $\Delta t_{\rm br}^{\rm iso}$. Light blue and light pink shading indicate the branches referred to
operationally as stable and unstable, respectively, over the simulated
time interval.}
\label{fig:isolated_clocks}
\end{figure}
%------------------------------------------------
\begin{figure}[t]
\centering
\includegraphics[width=\columnwidth]{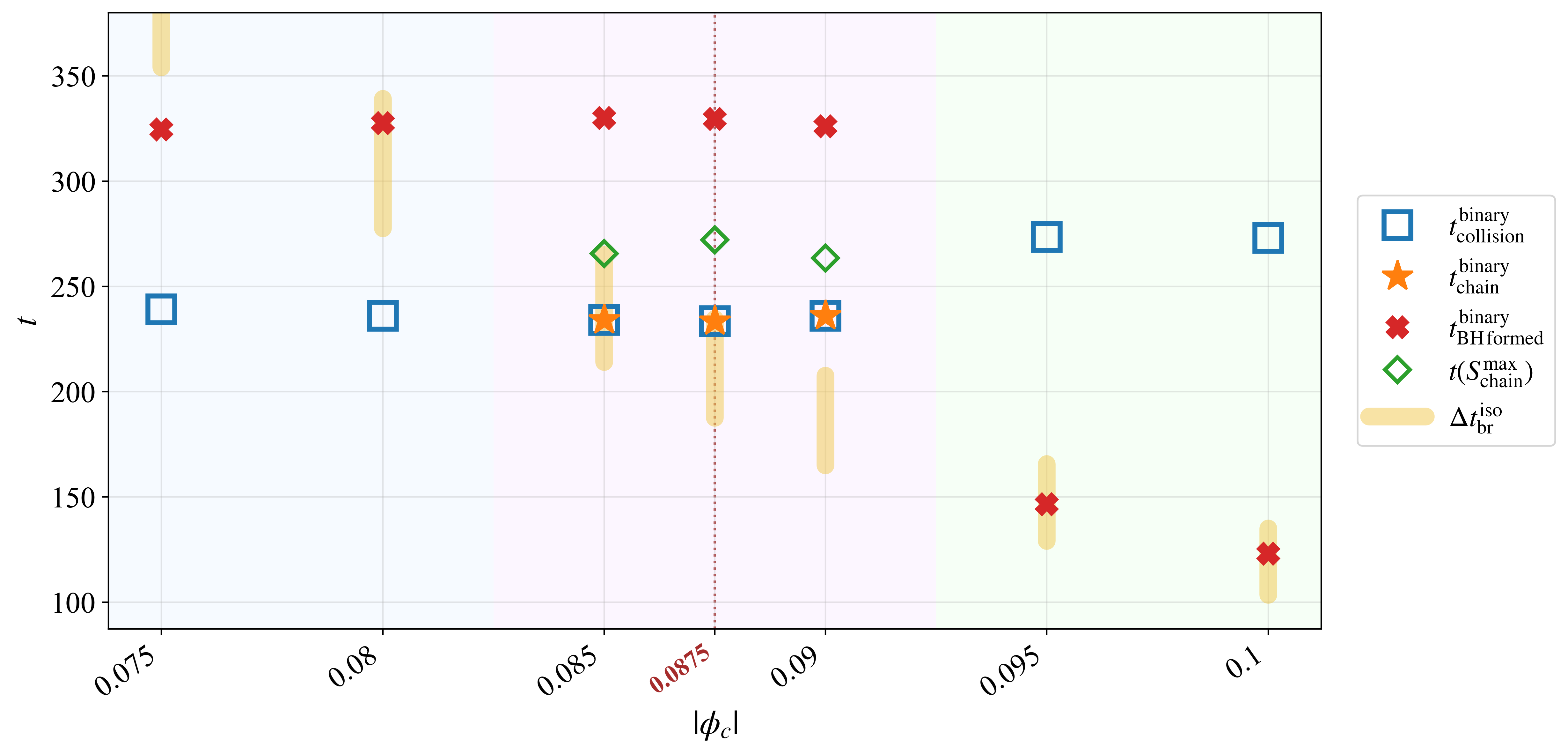}
\caption{
Characteristic times in binary head-on collisions at fixed $D=80$.
Yellow bars show the isolated breathing interval, blue squares mark the
binary collision time, orange stars mark the visible-chain onset, red
crosses mark the BH-formation time in the binary system, and green
diamonds mark the time at which the chain score reaches its maximum.
Visible chains appear only when the binary collision time is compatible
with the isolated breathing clock.
}
\label{fig:binary_timescale_scan}
\end{figure}
%------------------------------------------------

Figure~\ref{fig:isolated_clocks} first establishes the isolated breathing
clock for the \(n=2\), \(\lambda=400\) sequence. Our previous study of
massive boson stars showed that the stability change is organized by the
first maximum of the mass curve \(M(\phicen)\)~\cite{Ge:2025btw}. The
excited sequence considered here shows the same turning-point pattern.
The branch before the first maximum, \(dM/d|\phi_c|>0\), shows no
detectable instability over the simulated interval \(t\leq2000\),
with no sustained breathing stage or BH formation. After the first
maximum, the branch with \(dM/d|\phi_c|<0\) develops a finite
breathing stage followed by BH formation. Here the terms ``stable''
and ``unstable'' are used only as operational labels for the
finite-time behavior observed in the present axisymmetric evolutions,
and are not intended as a claim of generic linear stability of the
excited configurations in full \(3+1\) dimensions.

Recent studies have shown that repulsive self-interactions can lead
to a stability structure for excited boson stars that is more intricate
than a single turning-point criterion, both in relativistic evolutions
restricted to spherical dynamics and in the nonrelativistic linear
regime with generic perturbations~\cite{Sanchis-Gual:2021phr,
Nambo:2024gvs}. We have not attempted to map the present
relativistic \(n=2\) configurations onto the stability regions identified
in those works, since the excitation number, dynamical regime, and
perturbation sector are different. Accordingly, the ``stable'' and
``unstable'' labels used here refer only to the finite-time behavior
observed in the present axisymmetric evolutions, rather than to a global
mode-stability classification. For the timing-window analysis, the
relevant question is whether the internal instability becomes manifest
before the binary collision time.

The isolated sequence is sampled at
\(\phicen=0.01,0.03,\ldots,0.19\). Figure~\ref{fig:isolated_clocks}
shows the resulting trend: the orange bars mark the isolated breathing
window defined in Eq.~\eqref{eq:breathing_window}, and the red points
mark the BH formation time \(t_{\rm BH}^{\rm iso}\), identified by
\(\chi_{\min}<10^{-3}\). The detailed values of the breathing intervals
and BH times are listed in Table~S1 of the Supplemental Material. Along the unstable branch, the breathing window moves rapidly to earlier times and becomes shorter as \(\phicen\) increases. This provides the reference clock against which the binary collision time is compared.

We now compare this isolated clock with equal-mass, in-phase head-on
collisions of selected configurations on the unstable branch, evolved
with the same binary parameters \(D=80\) and \(v=0.1\). The purpose of
this comparison is not to treat \(\phicen\) as a continuous control
parameter for a single star, but to compare distinct equilibrium
configurations whose isolated evolutions have different breathing
windows. Figure~\ref{fig:binary_timescale_scan} overlays these isolated
breathing windows with the characteristic times extracted from the
corresponding binary evolutions: the collision time, the chain onset
time, the BH formation time, and the time \(t(S_{\rm chain}^{\max})\) at
which the chain score is maximal.

The comparison reveals a finite timing window for visible-chain
formation. For the lower-amplitude unstable cases, such as
\(\phicen=0.075\) and \(0.080\), the binary collision occurs before the
corresponding breathing window is reached, and no visible chain is
formed. For the intermediate cases,
\(\phicen=0.085\), \(0.0875\), and \(0.090\), the collision time lies
within, or sufficiently close to, the corresponding breathing window, and
visible chain-like transients are produced. A representative example for
\(\phicen=0.0875\) is shown in Fig.~\ref{fig:chain_snapshot}. For the
larger amplitudes shown here, \(\phicen=0.095\) and \(0.100\), the binary
system reaches the BH threshold before the nominal collision time.
Thus the subsequent evolution corresponds to a BH--BH merger, so no visible chain can form. 

In the visible-chain cases, the chain onset occurs at the binary
collision time,
\begin{equation}
t_{\rm chain}^{\rm binary}
\simeq
t_{\rm collision}^{\rm binary}.
\label{eq:tchain_tcollision}
\end{equation}
This coincidence links the chain-like morphology to the collision itself,
rather than to a late-time accidental deformation of the remnant. 

The comparison also shows why the isolated breathing window should be
viewed as a reference clock, not as a strict prediction of the binary
evolution. 
For \(\phicen=0.090\), the corresponding isolated evolution would have
reached the BH diagnostic threshold before the nominal binary collision
time, but the binary nevertheless forms a visible chain at collision and collapses only after
a finite post-chain interval. Conversely, for the larger amplitudes
\(\phicen=0.095\) and \(0.100\), the binary BH time can occur earlier
than the corresponding isolated BH time. These examples show that the
binary environment can shift the detailed timing relative to the isolated
evolution. The relevant criterion is therefore not a rigid isolated-time
boundary, but the compatibility between the binary collision time and the
isolated breathing clock.

The comparison at fixed \(D=80\) and \(v=0.1\) therefore supports the
timing-window interpretation. Radial excitation alone does not guarantee
visible chain formation. Instead, for the selected unstable
configurations studied here, the outcome depends on whether the binary
collision time is compatible with the breathing window identified from
the corresponding isolated evolution. This motivates the direct timing
test in the next section, where we fix \(\phicen=0.0875\) and vary the
initial separation \(D\) to shift the collision time.

%================================================
\section{Separation dependence of the timing window}
\label{sec:separation_scan}
%================================================
%------------------------------------------------
\begin{figure}[t]
\centering
\includegraphics[width=\columnwidth]{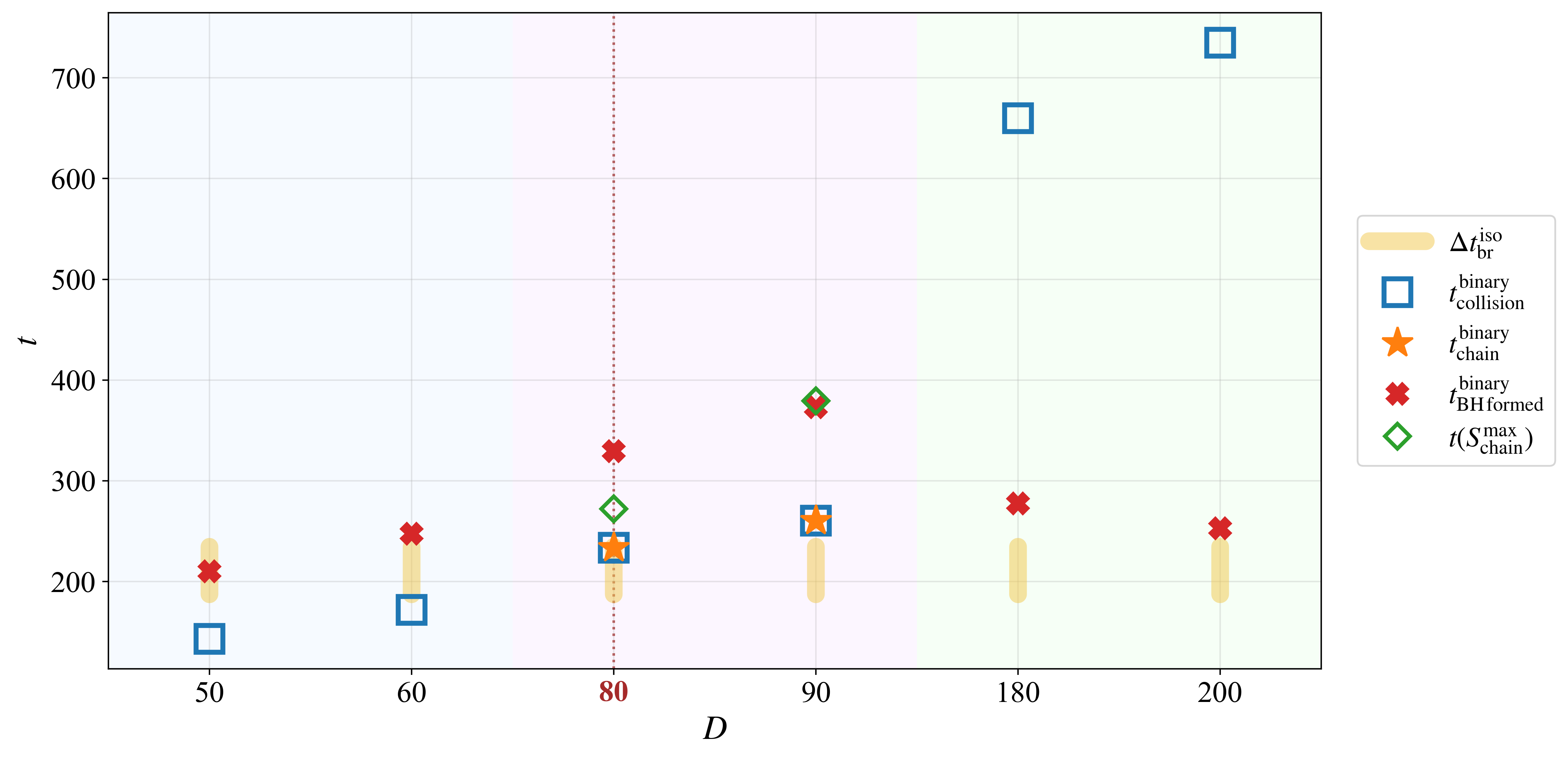}
\caption{
Separation test at fixed $\phicen=0.0875$. Varying $D$ shifts the
collision time relative to the same isolated breathing window. Early
collisions do not form visible chains, intermediate collisions do, and
large-separation cases form BHs before the nominal collision time,
preventing chain formation.
}
\label{fig:varyD_timing}
\end{figure}
%------------------------------------------------

We now test the timing interpretation by fixing the selected boson-star
configuration and varying the initial separation \(D\). We choose
\(\phicen=0.0875\), keep the initial velocity fixed at \(v=0.1\), and
compare six representative separations,
$
D=50,\ 60,\ 80,\ 90,\ 180,\ 200 .
$
The purpose of this comparison is to shift the binary collision time
relative to the isolated breathing window identified for the same
selected configuration. The result is shown in
Fig.~\ref{fig:varyD_timing}.

For \(D=50\) and \(D=60\), the binary collision occurs before the
isolated breathing window is reached. No visible chain is formed in
these cases.

For \(D=80\) and \(D=90\), the collision time lies within, or sufficiently
close to, the isolated breathing window. These are the cases in which
visible chain-like transients are produced. As in the \(\phicen\)
comparison, the chain onset occurs at the collision time, linking the
visible chain to the binary collision rather than to a late-time
accidental deformation.

For the larger separations, \(D=180\) and \(D=200\), the binary system
reaches the BH threshold before the nominal collision time. Thus the
subsequent evolution corresponds to a BH--BH merger, and no visible chain
can form.

The separation comparison therefore supports the timing-window
interpretation without changing the selected isolated boson-star
configuration. When the collision occurs too early relative to the
breathing window, no visible chain forms; when it occurs within, or close
to, the window, a visible chain appears; and when the system reaches the
BH threshold before the nominal collision time, the later evolution no
longer produces a chain. This provides a direct timing test of the
interpretation inferred from the \(\phicen\) comparison.

%================================================
\section{Discussion and outlook}
\label{sec:outlook}
%================================================

The simulations presented above support a timing-window interpretation
of visible-chain formation in collisions of radially excited boson stars.
Radial nodes are necessary for the internal structure considered here,
but they are not sufficient to guarantee a visible chain. For the selected
\(n=2\), \(\lambda=400\) configurations, visible chains appear only when
the binary collision time is compatible with the breathing window
identified from the corresponding isolated evolution. Collisions that
occur too early do not produce a visible chain, while configurations that
reach the BH threshold before the nominal collision time subsequently
undergo a BH--BH merger and cannot form a bosonic chain-like transient.

An important point is that the isolated breathing window should be viewed
as a reference clock, not as a strict prediction of the binary evolution.
The case \(\phicen=0.090\) shows that a binary can form a visible chain
even though the corresponding isolated evolution would have reached the
BH threshold before the nominal binary collision time.
Conversely,
for \(\phicen=0.095\) and \(0.100\), the BH formation time in the binary
can occur earlier than the corresponding isolated BH time. These examples
show that the binary environment can shift the detailed timing relative
to the isolated evolution. The robust statement is therefore not a rigid
inequality involving isolated times, but the compatibility between the
binary collision time and the breathing clock of the selected excited
configuration.

The breathing window should be understood as an operational proxy
for a transient dynamical state of the excited star, rather than as the
direct cause of chain formation. The \(n=2\) configuration contains a
nodeful radial profile with several scalar-field domains. In the present
analysis, the breathing criterion identifies a finite interval of
large-amplitude scalar-profile evolution during which the configuration
remains a coherent bosonic object. We interpret this interval as a stage
in which the internal shell structure may be particularly susceptible to
collision-induced reorganization. The present diagnostics do not directly
resolve the trajectories of the individual radial nodes.

The collision provides a second ingredient: an anisotropic external
perturbation with a preferred direction. After subtracting the
centre-of-mass acceleration, the leading differential effect of the
companion is tidal, stretching the scalar configuration along the
collision axis and compressing it in the transverse directions. When
strong overlap occurs while the node-supported radial structure is
undergoing this active redistribution, the collision-induced anisotropy
can reorganize the pre-existing shell structure into several
axis-supported density concentrations. In this interpretation, the
visible chain is a short-lived nonlinear morphology produced by the
coupling between internal radial restructuring and collision-induced
axial deformation, rather than a new equilibrium configuration.

This picture also provides a physical interpretation of the three
timing regimes observed in the simulations. If the collision occurs too
early, the internal restructuring has not developed sufficiently and no
resolved chain is produced. If strong overlap occurs during the active
restructuring stage, several axial density concentrations can form. If
the system reaches the BH diagnostic before common-remnant formation,
the coherent bosonic shell structure is no longer available to produce
a visible chain. The isolated breathing window therefore indicates the
stage during which the excited configuration is most susceptible to
collision-induced anisotropic reorganization.

The binary interaction can also modify the internal clock itself.
The delayed collapse of the \(\phicen=0.090\) binary relative to its
isolated evolution is consistent with such an effect. 
In addition, the
auxiliary BS--BH external-field evolution reported in the Supplemental
Material, constructed using the initial-data prescription of
Ref.~\cite{Marks:2026xvo}, is consistent with the possibility that an
external companion field can prolong or modify the breathing/tidally
distorted stage.
Taken together, these observations support the timing-window mechanism
at the dynamical level relevant to the present study: visible chain-like
transients arise only when the collision occurs during, or sufficiently
close to, the finite interval of active internal restructuring. The
present analysis does not, however, identify a unique linear eigenmode or
provide a complete mode-by-mode decomposition of the nonlinear coupling
responsible for this behavior. A more quantitative characterization would
require measurements of the local tidal tensor, binary node trajectories,
component-wise compactness, and their correlations with the shape tensor
\(\mathcal{Q}_{ij}\). Systematic BS--BH external-field evolutions and
phase-controlled BS--BS collisions would further help separate
companion-induced tidal deformation from phase-dependent scalar
interference and determine how each effect modifies the restructuring
and collapse time scales.

The main timing analysis focuses on equal-mass, in-phase, head-on
collisions of selected \(n=2\), \(\lambda=400\) configurations. An
additional fixed-separation check at \(\lambda=500\), reported in the
Supplemental Material, shows the same qualitative event ordering.
In the convention of Eq.~\eqref{eq:quartic_potential}, both choices
correspond to repulsive quartic self-interactions, \(\lambda>0\).

The present results should therefore not be assumed to carry over
unchanged to attractive self-interactions. The general organizing
idea---comparing the binary collision time with an intrinsic transient
restructuring time scale of the corresponding isolated configuration---
may remain relevant whenever such a time scale exists. However, changing
the sign of the self-interaction can modify the equilibrium sequence,
compactness, stability properties, and collapse time scales. It may
therefore shift, broaden, narrow, or even remove the timing interval
compatible with visible-chain formation, and may also change the
resulting morphology. Moreover, for the quartic-only potential adopted
here, taking \(\lambda<0\) makes the potential unbounded from below at
sufficiently large field amplitude. A physically complete attractive
model would therefore generally require additional stabilizing
higher-order terms or a different scalar potential.

The astrophysical formation of binaries containing radially excited
boson stars remains uncertain and is not addressed in the present work.
Previous spherically symmetric simulations nevertheless provide a proof
of principle that, for sufficiently strong repulsive self-interactions,
dilute nodeful scalar-field configurations can contract through
gravitational cooling while retaining an excited-state structure
~\cite{Sanchis-Gual:2021phr}. This evidence concerns the first excited
state under restricted symmetry assumptions and does not establish a
realistic formation channel for the \(n=2\) binaries considered here.
Long-term stability is not required for the present scenario: the
relevant condition is that the instability time scale be longer than
the binary collision time. The binaries studied here should therefore
be regarded as idealized controlled systems designed to isolate this
competition of time scales, rather than as a model of a generic
astrophysical binary population.

The extension to binaries with nonzero orbital angular momentum
depends on the dynamical regime. For a prompt off-axis encounter with
a finite impact parameter, closest approach, strong interaction, and
possible merger may still occur over a relatively short time scale. In
that case, the comparison between the intrinsic restructuring clock of
the isolated star and the time of closest approach or strong overlap
may remain a useful reference, although the resulting morphology need
not be an axis-aligned chain.

For a long inspiral, the same underlying competition of time scales
may remain relevant, but the appropriate internal clock need not coincide
with the breathing window measured from an isolated evolution. Each star
is exposed to the companion field over many orbital cycles, allowing
tidal deformation and mode excitation to accumulate before merger.
These effects may continuously modify the restructuring time scale of
the individual star. The timing-window picture would then need to be
generalized by comparing the merger or strong-interaction time with this
interaction-modified internal clock. The response is expected to depend
on the excitation number, compactness, and self-interaction of the
individual boson stars.

The present modified-Cartoon implementation assumes exact
axisymmetry, corresponding to an \(SO(2)\) rotational symmetry, and
therefore cannot represent either finite-impact-parameter encounters or
inspiralling binaries. Testing the generalized timing interpretation in
these regimes requires fully three-dimensional, nonaxisymmetric
simulations and is left for future work. Within the present axisymmetric
framework, broader surveys in excitation number, repulsive
self-interaction strength, relative phase, and head-on boost, together
with studies of stabilized attractive potentials, can further test the
robustness of the timing-window mechanism. Fully three-dimensional
studies can subsequently extend this program to finite-impact-parameter
encounters and inspiralling binaries, while a unified comparison of
chain and ring channels provides an additional direction for assessing
its generality.

%=================================================================
\begin{acknowledgments}
The author acknowledges HIAS for access to the ``Quantum Universe
Physical Simulation Platform''. This work used the DiRAC Memory
Intensive service (COSMA8) at Durham University, managed by the
Institute for Computational Cosmology on behalf of the STFC DiRAC HPC
Facility (www.dirac.ac.uk). The DiRAC service at Durham was funded by
BEIS, UKRI and STFC capital funding, Durham University and STFC
operations grants. DiRAC is part of the UKRI Digital Research
Infrastructure. This work was supported by the National Natural Science
Foundation of China under Grant No.~12505066.
\end{acknowledgments}

\clearpage
\bibliography{BS}

\end{document}

\title{Supplemental Material for ``Timing-Window Mechanism for Chain-Like Transients in Collisions of Radially Excited Boson Stars''}

\author{Bo-Xuan Ge
\orcidlink{0000-0003-0738-3473}}
\email{bo-xuan.ge@ucas.ac.cn}
\affiliation{\hias}
\affiliation{\kcl}

\maketitle

%==================================================================
\section{Numerical setup}
\label{sec:supp_numerical_setup}
%==================================================================

The simulations are performed with our {\sc GRChombo}-based numerical-relativity
code using the CCZ4 formulation and adaptive mesh refinement.  The systems
considered in the main text are axisymmetric: isolated stars are evolved
with the same axisymmetric infrastructure, and the binaries are equal-mass,
in-phase, zero-impact-parameter head-on collisions.  We therefore use the
modified Cartoon method to evolve the corresponding three-dimensional
axisymmetric dynamics on a two-dimensional computational domain.  The
general Cartoon construction follows Refs.~\cite{Cook:2016soy,
Cook:2016qnt}, and its implementation for boson-star evolutions follows
Refs.~\cite{Ge:2024fum,Ge:2025btw,Ge:2024itl}.

The computational domain has length \(512\) in code units.  Throughout this
Supplemental Material, \(N\) denotes the number of grid points on the
coarsest AMR level covering the full domain.  Thus the coarsest-level grid
spacing is
\begin{equation}
\Delta x_0=\frac{512}{N}.
\label{eq:supp_dx0_from_N}
\end{equation}
For the fiducial simulations we use \(N=256\), giving
\begin{equation}
\Delta x_0=2 .
\label{eq:supp_dx0}
\end{equation}
The adaptive mesh has refinement levels \(\ell=0,\ldots,6\), and the
spacing on level \(\ell\) is
\begin{equation}
\Delta x_\ell=\frac{\Delta x_0}{2^\ell},
\qquad
\ell=0,\ldots,6 .
\label{eq:supp_dx_levels}
\end{equation}
Therefore the finest-grid spacing in the fiducial simulations is
\begin{equation}
\Delta x_6=\frac{2}{2^6}=\frac{1}{32}.
\label{eq:supp_dx_finest}
\end{equation}
Unless otherwise stated, the quoted event times in the main text correspond
to the fiducial coarsest-level resolution \(N=256\).  In the resolution
check below, changing \(N\) changes the coarsest-level spacing according to
Eq.~\eqref{eq:supp_dx0_from_N}, while keeping the same refinement hierarchy.

Unless otherwise stated, configurations use the quartic self-interaction
potential in the main text, with \(m=1\), \(\lambda=400\), and radial
excitation number \(n=2\).  The additional self-interaction check below
uses the same setup with \(\lambda=500\). The isolated equilibrium stars are obtained by solving the boson-star
shooting problem with the harmonic ansatz
\begin{equation}
\phi(t,r)=|\phi(r)|e^{i\omega t},
\label{eq:supp_bs_ansatz}
\end{equation}
following the construction used in Ref.~\cite{Helfer:2021brt}.  Binary
initial data are constructed from two equal-mass, in-phase stars using the
improved superposition method of
Refs.~\cite{Helfer:2021brt,Helfer:2018vtq}.  In the fixed-separation scan
we set
\begin{equation}
D=80,
\qquad
v=0.1,
\label{eq:supp_fiducial_binary}
\end{equation}
and vary \(\phicen\).  In the direct separation test we fix
\begin{equation}
\phicen=0.0875,
\qquad
v=0.1,
\label{eq:supp_varyD_fixed}
\end{equation}
and use the representative separations
\begin{equation}
D=50,\ 60,\ 80,\ 90,\ 180,\ 200 .
\label{eq:supp_varyD_values}
\end{equation}

Throughout the timing analysis, ``BH formation'' is used in the same
operational sense as in the main text.  We define the diagnostic BH time as
the first output time at which
\begin{equation}
\chi_{\min}<10^{-3},
\label{eq:supp_chi_threshold}
\end{equation}
where \(\chi_{\min}\) is the minimum of the conformal factor on the
computational domain.

%==================================================================
\section{Diagnostics and event-time extraction}
\label{sec:supp_diagnostics}
%==================================================================

This section defines the diagnostics used to extract the event times shown
in the main text.  The aim is not to introduce a new invariant observable,
but to apply the same operational criteria to all runs in the timing-window
comparison.

%----------------------------------------------------------
\subsection{Isolated-star clocks}
\label{sec:supp_isolated_clocks_def}
%----------------------------------------------------------

For each isolated star we extract the isolated reference clock from the
time series of the maximum scalar-field amplitude and the conformal-factor
diagnostic.  At each output time we compute
\begin{equation}
\Phi_{\max}(t)=\max_{\mathbf{x}}|\phi(t,\mathbf{x})|,
\label{eq:supp_phimax}
\end{equation}
and
\begin{equation}
\chi_{\min}(t)=\min_{\mathbf{x}}\chi(t,\mathbf{x}) .
\label{eq:supp_chimin}
\end{equation}

The breathing window is identified from the relative change of
\(\Phi_{\max}(t)\) with respect to its first valid output value.  We define
\begin{equation}
\mathcal{R}_{\Phi}(t)
=
\frac{\Phi_{\max}(t)}{\Phi_{\max}(t_0)},
\label{eq:supp_phi_ratio}
\end{equation}
where \(t_0\) denotes the first valid diagnostic output time.  The start
time \(\tbrstartiso\) is defined as the first output time at which
\begin{equation}
\mathcal{R}_{\Phi}(t)\geq 1.03
\label{eq:supp_breath_start_ratio}
\end{equation}
is sustained for three consecutive diagnostic outputs.  The end time
\(\tbrendiso\) is defined as the first subsequent output time, after
\(\tbrstartiso\), at which
\begin{equation}
\mathcal{R}_{\Phi}(t)\leq 1.00
\label{eq:supp_breath_end_ratio}
\end{equation}
is sustained for three consecutive diagnostic outputs.  The corresponding
duration is
\begin{equation}
\dtbriso=\tbrendiso-\tbrstartiso .
\label{eq:supp_breath_duration}
\end{equation}
If either the start or the end condition is not met within the simulated
time interval, the corresponding entry is recorded as a dash in the tables.

As an auxiliary diagnostic of internal restructuring, we also monitor the
first sustained displacement of the node-supported scalar-field profile.
This diagnostic is applied to two classes of quantities extracted from
symmetry-axis lineouts.  The first class consists of the positions of
resolved radial nodes,
\begin{equation}
x_i(t),\qquad y_i(t),
\label{eq:supp_node_positions}
\end{equation}
where \(i\) labels the resolved nodes along the two lineouts.  The second
class consists of the separations between pairs of resolved radial nodes,
\begin{equation}
x_{ij}(t)=|x_i(t)-x_j(t)|,
\qquad
y_{ij}(t)=|y_i(t)-y_j(t)| .
\label{eq:supp_node_separations}
\end{equation}
For either a node position or a node separation, denoted collectively by
\(q(t)\), a sustained node shift is counted when
\begin{equation}
|q(t)-q(t_0)|
\geq
\max\left(1.0,\ 0.10\,|q(t_0)|\right)
\label{eq:supp_node_shift_threshold}
\end{equation}
is sustained for three consecutive diagnostic outputs.  Here \(t_0\) again
denotes the first valid diagnostic output time for the corresponding
quantity.  The absolute floor of one code-length unit prevents small
numerical displacements from being identified as node motion, while the
relative threshold requires a ten-percent change when the initial node
position or node separation is large.

This node-shift diagnostic is introduced only as an auxiliary measure
for characterizing possible restructuring of the node-supported profile
during the scalar-amplitude breathing window.
The breathing-window times quoted in the tables are
determined by Eqs.~\eqref{eq:supp_breath_start_ratio} and
\eqref{eq:supp_breath_end_ratio}.

The isolated BH time \(\tBHiso\) is defined as the first output time at
which
\begin{equation}
\chi_{\min}(t)< 10^{-3}.
\end{equation}
A dash in the tables below means that the corresponding breathing or BH
event is not detected within the simulated interval.

%----------------------------------------------------------
\subsection{Bridge diagnostic and binary collision time}
\label{sec:supp_bridge}
%----------------------------------------------------------

For a binary run, the collision time is defined by the formation of a
common bosonic remnant.  We determine this time using a bridge diagnostic
constructed from the density weight close to the collision axis, and we
monitor the density-weighted separation of the two objects as an auxiliary
check.  The bridge diagnostic measures the filling of the central region
between the two initially separated stars, while the separation diagnostic
checks that the two density-weighted cores approach each other consistently.

Let \(x\) denote the collision-axis coordinate, and let \(y\geq0\) denote
the transverse cylindrical radius in the axisymmetric diagnostic plane.
At each output time we construct a one-dimensional near-axis density-weight
profile along the collision axis.  Let \(X_i\) be the centre of the \(i\)-th
axial sampling interval, and let \(\Delta X\) be the corresponding
diagnostic sampling spacing.  We define
\begin{equation}
P_x(X_i,t)
=
\int_{X_i-\Delta X/2}^{X_i+\Delta X/2}\du x
\int_{0}^{R_{\rm ax}}\du y\,
\rho(t,x,y)\sqrt{\gamma(t,x,y)}\,2\pi y .
\label{eq:supp_axis_profile}
\end{equation}
In the numerical diagnostic this expression is evaluated as a discrete sum
over the sampled grid points.  The factor \(2\pi y\) is the cylindrical
volume factor appropriate to the axisymmetric reduction.

The parameter \(R_{\rm ax}\) defines the transverse radius of the near-axis
region used to construct the bridge diagnostic.  It is not the transverse
size of the full computational domain.  Instead, we choose
\begin{equation}
R_{\rm ax}=f_{\rm axis}R_{\rm samp},
\qquad
f_{\rm axis}=0.10 ,
\label{eq:supp_axis_region}
\end{equation}
where \(R_{\rm samp}\) is the characteristic size of the diagnostic sampling
region.  Thus \(P_x(X_i,t)\) measures the density weight close to the
collision axis as a function of \(x\).  Restricting the integral to
\(0\leq y\leq R_{\rm ax}\) makes the bridge diagnostic sensitive to the
filling of the central region between the two stars, rather than to
off-axis material.

Let \(x_{\rm c}\) be the midpoint between the two initially separated
stars, and let \(i_{\rm c}\) label the axial sampling interval whose centre
\(X_i\) is closest to \(x_{\rm c}\).  The bridge value is defined as the
average near-axis density weight over the central sampling intervals around
this midpoint,
\begin{equation}
P_{\rm br}(t)
=
\frac{1}{2n_{\rm br}+1}
\sum_{k=-n_{\rm br}}^{n_{\rm br}}
P_x(X_{i_{\rm c}+k},t),
\qquad
n_{\rm br}=3 .
\label{eq:supp_bridge_value}
\end{equation}
The integer \(n_{\rm br}\) is the half-width of the central bridge region
measured in axial sampling intervals.  Thus \(n_{\rm br}=3\) means that the
bridge value is averaged over seven central sampling intervals.  This
choice matches the minimum separation used in the peak-finding diagnostic
and prevents a single noisy sampling interval from determining the bridge
ratio.

The left and right object peaks are defined from the same near-axis
profile, excluding the central bridge region,
\begin{equation}
P_{\rm L}(t)
=
\max_{i<i_{\rm c}-n_{\rm br}} P_x(X_i,t),
\qquad
P_{\rm R}(t)
=
\max_{i>i_{\rm c}+n_{\rm br}} P_x(X_i,t).
\label{eq:supp_bridge_peaks}
\end{equation}
The bridge ratio is then defined by
\begin{equation}
B(t)
=
\frac{P_{\rm br}(t)}
{\min\left[P_{\rm L}(t),P_{\rm R}(t)\right]} .
\label{eq:supp_bridge_ratio}
\end{equation}
The denominator uses the weaker of the two object peaks so that the
diagnostic is not biased by a mild left-right asymmetry.  Before merger,
the central region between the two stars is depleted and \(B(t)\ll1\).
As the two configurations form a common remnant, the central region fills
in and \(B(t)\) increases.

In addition to \(B(t)\), we monitor the density-weighted separation between
the two objects.  This is computed from the full diagnostic density weight
rather than only from the near-axis profile.  We define
\begin{equation}
\du W
=
\rho(t,x,y)\sqrt{\gamma(t,x,y)}\,2\pi y\,\du x\,\du y .
\label{eq:supp_density_weight_binary}
\end{equation}
The left and right density-weighted centres are
\begin{equation}
x_{\rm L}(t)
=
\frac{
\int_{x<x_{\rm c}} x\,\du W
}{
\int_{x<x_{\rm c}} \du W
},
\qquad
x_{\rm R}(t)
=
\frac{
\int_{x>x_{\rm c}} x\,\du W
}{
\int_{x>x_{\rm c}} \du W
}.
\label{eq:supp_binary_centres}
\end{equation}
The corresponding separation diagnostic is
\begin{equation}
d_{\rm cm}(t)=x_{\rm R}(t)-x_{\rm L}(t).
\label{eq:supp_binary_separation}
\end{equation}
This quantity provides an auxiliary check that the two objects are
approaching and that the bridge-ratio crossing occurs when the two
density-weighted cores have become close.  We do not use
\(d_{\rm cm}(t)\) itself as the primary definition of the collision time,
because after common-remnant formation the division into left and right
halves is only a diagnostic convention rather than a physical separation
between two independent stars.

We use three operational bridge levels,
\begin{equation}
B_{\rm contact}=0.10,
\qquad
B_{\rm common}=0.35,
\qquad
B_{\rm strong}=0.50 .
\label{eq:supp_bridge_thresholds}
\end{equation}
These numbers are empirical diagnostic thresholds, not physical constants.
They quantify the filling of the central region relative to the weaker of
the two object peaks.  The contact level records the first weak filling of
the region between the two stars, and the strong level records a more
developed bridge.  The timing analysis uses only the intermediate level
\(B_{\rm common}=0.35\), which gives a uniform operational definition of
common-remnant formation across the scan.  Its role is to fix a consistent
timing convention, not to define a sharp physical phase transition.

The collision time quoted in the main text is therefore
\begin{equation}
\tcollbin
=
\min\left\{
t:\ B(t)\geq B_{\rm common}
\ \text{for three consecutive diagnostic outputs}
\right\}.
\label{eq:supp_collision_time}
\end{equation}
The three-output sustainment condition prevents a transient fluctuation of
the central density profile from being misidentified as common-remnant
formation.  In all cases used in the timing-window comparison, this bridge
criterion is checked against the simultaneous decrease of
\(d_{\rm cm}(t)\), ensuring that the assigned collision time corresponds to
the approach and connection of the two objects rather than to an isolated
density fluctuation near the midpoint.

%----------------------------------------------------------
\subsection{Morphology scores and labels}
\label{sec:supp_morphology}
%----------------------------------------------------------

After common-remnant formation we characterize the remnant morphology using
density-weighted diagnostics constructed on the same axisymmetric sampling
region as the bridge diagnostic.  The purpose of these quantities is not to
define a gauge-invariant observable, but to provide a uniform operational
criterion for identifying chain-like remnants across the parameter scan.

We use the density weight
\begin{equation}
\du W
=
\rho(t,x,y)\sqrt{\gamma(t,x,y)}\,2\pi y\,\du x\,\du y ,
\label{eq:supp_morph_weight}
\end{equation}
and the total sampled weight
\begin{equation}
W(t)
=
\int_{\mathcal{D}_{\rm samp}}\du W ,
\label{eq:supp_total_weight}
\end{equation}
where \(\mathcal{D}_{\rm samp}\) denotes the diagnostic sampling region.
Let \(R_{\rm samp}\) be the characteristic size of this sampling region,
defined by the sampled spatial extent rather than by the full computational
domain.

We split the sampled density weight into a near-axis part and an off-axis
part.  The near-axis region is defined by
\begin{equation}
\mathcal{D}_{\rm axis}
=
\left\{
(x,y)\in\mathcal{D}_{\rm samp}:
0\leq y\leq R_{\rm ax}
\right\},
\qquad
R_{\rm ax}=f_{\rm axis}R_{\rm samp},
\qquad
f_{\rm axis}=0.10 .
\label{eq:supp_axis_region_morph}
\end{equation}
The off-axis region is defined by
\begin{equation}
\mathcal{D}_{\rm off}
=
\left\{
(x,y)\in\mathcal{D}_{\rm samp}:
y\geq R_{\rm off}
\right\},
\qquad
R_{\rm off}=f_{\rm off}R_{\rm samp},
\qquad
f_{\rm off}=0.20 .
\label{eq:supp_off_axis_region_morph}
\end{equation}
The corresponding density-weight fractions are
\begin{equation}
F_{\rm axis}(t)
=
\frac{1}{W(t)}
\int_{\mathcal{D}_{\rm axis}}\du W ,
\qquad
F_{\rm off}(t)
=
\frac{1}{W(t)}
\int_{\mathcal{D}_{\rm off}}\du W .
\label{eq:supp_axis_off_fractions}
\end{equation}
Thus \(F_{\rm axis}\) measures the fraction of density weight supported
close to the collision axis, while \(F_{\rm off}\) measures the competing
off-axis contribution.

The near-axis and off-axis regions are not complementary.  The intermediate
buffer region
\begin{equation}
\mathcal{D}_{\rm mid}
=
\left\{
(x,y)\in\mathcal{D}_{\rm samp}:
R_{\rm ax}<y<R_{\rm off}
\right\}
\label{eq:supp_mid_region_morph}
\end{equation}
is included in the total weight \(W(t)\), but is not included in either
\(F_{\rm axis}\) or \(F_{\rm off}\).  This buffer region prevents matter
near the axis/off-axis boundary from being assigned sharply to either
category.  Material in \(\mathcal{D}_{\rm mid}\) therefore does not directly
support the chain score through \(F_{\rm axis}\), nor does it directly enter
the off-axis penalty through \(F_{\rm off}\), although it reduces both
fractions through the common normalization by \(W(t)\).

As an auxiliary shape diagnostic, we also record the axis-depletion measure
\begin{equation}
D_{\rm axis}(t)
=
1-
\frac{\rho_{\max}^{\rm axis}(t)}{\rho_{\max}(t)} ,
\label{eq:supp_axis_depletion}
\end{equation}
where \(\rho_{\max}^{\rm axis}\) is the maximum density inside
\(\mathcal{D}_{\rm axis}\), and \(\rho_{\max}\) is the maximum density in
\(\mathcal{D}_{\rm samp}\).  This quantity records whether the maximum
density has moved away from the near-axis region.  It is not used by itself
to define a visible chain.

The near-axis density-weight profile \(P_x(X_i,t)\), defined in
Eq.~\eqref{eq:supp_axis_profile}, is used to count significant
axis-supported peaks.  We first apply a three-point smoothing,
\begin{equation}
\widetilde{P}_x(X_i,t)
=
\frac{
P_x(X_{i-1},t)+P_x(X_i,t)+P_x(X_{i+1},t)
}{3}.
\label{eq:supp_smoothed_axis_profile}
\end{equation}
A sampling point \(X_i\) is counted as a candidate peak only if it is a
local maximum of the smoothed profile,
\begin{equation}
\widetilde{P}_x(X_i,t)>
\widetilde{P}_x(X_{i-1},t),
\qquad
\widetilde{P}_x(X_i,t)>
\widetilde{P}_x(X_{i+1},t).
\label{eq:supp_local_peak_condition}
\end{equation}
It is retained as a significant axis-supported peak only if its height
satisfies
\begin{equation}
\widetilde{P}_x(X_i,t)
\geq
\epsilon_{\rm peak}
\max_j \widetilde{P}_x(X_j,t),
\qquad
\epsilon_{\rm peak}=0.25 ,
\label{eq:supp_peak_threshold}
\end{equation}
and if neighbouring retained peaks are separated by at least
\begin{equation}
n_{\rm sep}=3
\label{eq:supp_peak_min_sep}
\end{equation}
axial sampling intervals.  We denote the resulting number of retained
local maxima by \(N_{\rm peak}^{x}(t)\).

The chain score is defined as
\begin{equation}
S_{\rm chain}(t)
=
\max\!\left(0,N_{\rm peak}^{x}(t)-2\right)
F_{\rm axis}(t)
\max\!\left[
0,
\frac{
F_{\rm axis}(t)-\kappa_{\rm off}F_{\rm off}(t)
}{
F_{\rm axis}(t)+\kappa_{\rm off}F_{\rm off}(t)+10^{-300}
}
\right],
\label{eq:supp_chain_score}
\end{equation}
with
\begin{equation}
\kappa_{\rm off}=0.75 .
\label{eq:supp_chain_offaxis_factor}
\end{equation}
This score combines three requirements.  The factor
\(\max(0,N_{\rm peak}^{x}-2)\) counts the number of significant
axis-supported peaks in excess of the two pre-merger object cores, so that
the original binary configuration is not itself counted as a chain.  The
factor \(F_{\rm axis}\) requires the density weight to be concentrated near
the collision axis.  The final factor is an axis-dominance factor: it is
positive only when the near-axis density weight exceeds the off-axis
contribution after weighting the latter by \(\kappa_{\rm off}\).  Therefore
off-axis material suppresses the chain score, while a clean multi-peak
near-axis structure gives a larger \(S_{\rm chain}\).

An instantaneous chain-like label is assigned only after the common-remnant
bridge condition has been reached.  Before common-remnant formation, the
system is treated as a pre-merger binary and no chain morphology is
assigned.  After common-remnant formation, a chain-like label requires
\begin{equation}
S_{\rm chain}(t)\geq S_{\rm chain}^{\rm lab},
\qquad
S_{\rm chain}^{\rm lab}=0.05,
\label{eq:supp_chain_label_threshold}
\end{equation}
together with
\begin{equation}
N_{\rm peak}^{x}(t)\geq3,
\qquad
F_{\rm axis}(t)\geq0.30,
\qquad
F_{\rm axis}(t)>0.75F_{\rm off}(t).
\label{eq:supp_chain_label_conditions}
\end{equation}
These conditions ensure that the label corresponds to a resolved multi-peak
structure supported along the collision axis, rather than to the two
pre-merger object cores, a weak residual feature, or off-axis material.

To avoid assigning a chain-like label to extremely weak late-time material,
we use a weak-tail guard.  Let
\begin{equation}
\rho_{\max}^{\rm rel}(t)
=
\frac{\rho_{\max}(t)}{\rho_{\max}(t_0)},
\qquad
W^{\rm rel}(t)
=
\frac{W(t)}{W(t_0)},
\label{eq:supp_weak_tail_relative}
\end{equation}
where \(t_0\) is the first valid morphology-output time.  If either
\begin{equation}
\rho_{\max}^{\rm rel}(t)<10^{-3}
\qquad
\text{or}
\qquad
W^{\rm rel}(t)<10^{-3},
\label{eq:supp_weak_tail_guard}
\end{equation}
the output is treated as a weak-tail state and no chain morphology is
assigned.

The chain onset time \(\tchainbin\) used in the timing comparison is not
defined by a single isolated occurrence of the instantaneous chain-like
label.  Instead, we require the chain-like label defined above to be
sustained for three consecutive diagnostic outputs.  Let \(t_i\) denote the
discrete morphology-output times.  We define
\begin{equation}
\tchainbin
=
\min_i
\left\{
t_i:
\text{the chain-like label is assigned at }
t_i,\ t_{i+1},\ \text{and } t_{i+2}
\right\}.
\label{eq:supp_chain_time}
\end{equation}
The three-output sustainment condition prevents a transient single-output
feature from being assigned as a chain-onset event.

Finally, the visible-chain classification used in the main text is more
restrictive than the instantaneous chain-like label. We define the valid
maximum chain score by taking the maximum only over outputs that remain
in the pre-BH classification channel,
\begin{equation}
S^{\max}_{\rm chain,valid}
=
\max_{t<t^{\rm binary}_{\rm BH\,formed}} S_{\rm chain}(t).
\end{equation}
A binary is counted as a visible-chain case only if a sustained
chain-containing label is present and
\begin{equation}
S^{\max}_{\rm chain,valid} \ge S_{\rm vis},
\qquad
S_{\rm vis}=2.0 .
\end{equation}
Raw morphology scores after the system has entered the BH channel are
not used as visible-chain evidence.

\clearpage
%==================================================================
\section{Isolated breathing windows}
\label{sec:supp_isolated_table}
%==================================================================

Table~\ref{tab:supp_isolated_times} lists a coarse isolated sequence
used to illustrate the trend of the breathing intervals and BH diagnostic
times along the \(n=2\), \(\lambda=400\) sequence.  The breathing
intervals are extracted from the relative-amplitude criterion in
Eqs.~\eqref{eq:supp_phi_ratio}--\eqref{eq:supp_breath_end_ratio}, and the
BH diagnostic time is extracted from the conformal-factor threshold in
Eq.~\eqref{eq:supp_chi_threshold}.  The fine isolated clocks used in the
fixed-separation and separation timing comparisons are included directly
in Tables~\ref{tab:supp_fixed_separation_summary}
and~\ref{tab:supp_separation_summary}.

For small central amplitudes no closed breathing interval or BH diagnostic
event is detected within the simulated time interval.  For larger central
amplitudes, the isolated configurations develop finite breathing intervals
followed by collapse according to the conformal-factor diagnostic.  The
breathing interval is used only as a reference clock for the binary problem;
it is not assumed to be a strict predictor of the binary evolution.

%-----------------------------------------------------
\begin{table}[t]
\centering
\caption{Isolated-star breathing intervals and BH diagnostic times for the
\(n=2\), \(\lambda=400\) sequence.  The breathing start and end times are
defined by the relative-amplitude criterion
\(\mathcal{R}_{\Phi}\geq1.03\) and the subsequent sustained return to
\(\mathcal{R}_{\Phi}\leq1.00\), respectively.  The BH time is defined by
\(\chi_{\min}<10^{-3}\).  A dash indicates that the corresponding event
is not detected within the simulated interval \(t\leq2000\).}
\label{tab:supp_isolated_times}
\begin{ruledtabular}
\begin{tabular}{ccccc}
\(\phicen\) &
\(\tbrstartiso\) &
\(\tbrendiso\) &
\(\dtbriso\) &
\(\tBHiso\) \\
\hline
0.01 & --    & --    & --   & --     \\
0.03 & --    & --    & --   & --     \\
0.05 & 1070  & 1134  & 64   & 1231.5 \\
0.07 & 451.5 & 541.5 & 90   & 555.5  \\
0.09 & 165   & 207.5 & 42.5 & 225.5  \\
0.11 & 70.5  & 95    & 24.5 & 117.5  \\
0.13 & 38    & 54.5  & 16.5 & 83     \\
0.15 & 22.5  & 34.5  & 12   & 70.5   \\
0.17 & 14    & 23    & 9    & 62.5   \\
0.19 & 8.5   & 16    & 7.5  & 46.5   \\
\end{tabular}
\end{ruledtabular}
\end{table}

\clearpage

%==================================================================
\section{Fixed-separation timing-window summary}
\label{sec:supp_eta_classifier}
%==================================================================

We summarize the fixed-separation amplitude scan by comparing the binary
collision time with the isolated breathing window of the corresponding
single star.  For each amplitude with a closed isolated breathing interval,
we define the normalized timing variable
\begin{equation}
\eta
=
\frac{\tcollbin-\tbrstartiso}
{\tbrendiso-\tbrstartiso} .
\label{eq:supp_eta}
\end{equation}
Thus \(0\leq\eta\leq1\) means that the binary collision time lies inside
the isolated breathing window, \(\eta<0\) means that the collision occurs
before the isolated breathing window, and \(\eta>1\) means that it occurs
after the end of that isolated interval.  This quantity is used only as a
compact timing diagnostic.  The isolated interval is a reference clock, not
a rigid boundary for the binary dynamics.

Table~\ref{tab:supp_fixed_separation_summary} lists the timing quantities used in the
fixed-separation comparison.  The column
\(\tBHbin-\tcollbin\) records whether the binary remains outside the BH
channel at the time of common-remnant formation.  Positive values mean that
the common remnant forms before the BH diagnostic threshold is reached,
while negative values mean that the system has already entered the BH channel before the nominal collision time.  The visible-chain
classification additionally requires a valid sustained chain-like morphology
before the BH diagnostic threshold and a maximum valid chain score satisfying
\begin{equation}
S_{\rm chain,valid}^{\max}\geq S_{\rm vis},
\qquad
S_{\rm vis}=2.0 .
\label{eq:supp_fixedD_visible_rule}
\end{equation}
Raw morphology scores after the system has entered the BH channel are not
used as visible-chain evidence.

%------------------------------------------------------
\begin{table*}[t]
\caption{Timing-window summary for the fixed-separation amplitude scan
with \(D=80\) and \(v=0.1\). The normalized timing variable \(\eta\)
compares the binary collision time with the corresponding isolated
breathing interval. The score column reports
\(S_{\rm chain,valid}^{\max}\), the maximum chain score in the valid
pre-BH classification channel. The visible-chain threshold is
\(S_{\rm vis}=2.0\). A dash indicates that the case is excluded by event
ordering because the BH diagnostic threshold is reached before
common-remnant formation.}
\begin{ruledtabular}
\begin{tabular}{cccccccl}
\(\phicen\) &
\(t_{\rm br,start}^{\rm iso}\) &
\(t_{\rm br,end}^{\rm iso}\) &
\(\eta\) &
\(t_{\rm collision}^{\rm binary}\) &
\(t_{\rm BH\,formed}^{\rm binary}
 - t_{\rm collision}^{\rm binary}\) &
\(S_{\rm chain,valid}^{\max}\) &
Classification \\
\hline
0.075  & 354.0 & 429.0 & \(-1.53\) & 239.0 & \(+85.5\)  & 0.999 & weak candidate \\
0.080  & 277.5 & 339.0 & \(-0.67\) & 236.0 & \(+91.5\)  & 0.999 & weak candidate \\
0.085  & 214.0 & 265.0 & \( 0.39\) & 234.0 & \(+96.0\)  & 2.961 & visible chain \\
0.0875 & 187.5 & 234.5 & \( 0.95\) & 232.0 & \(+97.5\)  & 2.898 & visible chain \\
0.090  & 165.0 & 207.5 & \( 1.67\) & 236.0 & \(+90.0\)  & 2.963 & visible chain \\
0.095  & 129.0 & 165.5 & \( 3.96\) & 273.5 & \(-127.0\) & --    & BH before collision \\
0.100  & 103.5 & 135.0 & \( 5.38\) & 273.0 & \(-150.0\) & --    & BH before collision \\
\end{tabular}
\end{ruledtabular}\label{tab:supp_fixed_separation_summary}
\end{table*}
%------------------------------------------------------

The first two cases collide before the corresponding isolated breathing
clock and produce only weak chain candidates, with
\(S_{\rm chain,valid}^{\max}<S_{\rm vis}\).  The intermediate amplitudes
produce visible chains, with
\(S_{\rm chain,valid}^{\max}\geq S_{\rm vis}\) and
\(\tBHbin-\tcollbin>0\).  The case \(\phicen=0.090\) illustrates why the
isolated clock should not be treated as a sharp cutoff: although
\(\eta>1\), the binary remains outside the BH channel through
common-remnant formation and forms a visible chain before collapse.  By
contrast, the \(\phicen=0.095\) and \(0.100\) binaries have
\(\tBHbin-\tcollbin<0\), so the BH diagnostic threshold is reached before
common-remnant formation.  These cases are therefore excluded by event
ordering and are not assigned a valid visible-chain score.

\clearpage
%==================================================================
\section{Separation comparison at fixed \texorpdfstring{\(\phicen=0.0875\)}{phic=0.0875}}
\label{sec:supp_varyD}
%==================================================================

The separation scan provides a direct timing test of the mechanism proposed
in the main text.  In contrast to the fixed-separation amplitude scan, here
the isolated star is kept fixed at \(\phicen=0.0875\).  Therefore the
isolated breathing window is the same for all entries,
\begin{equation}
\tbrstartiso=187.5,
\qquad
\tbrendiso=234.5,
\qquad
\dtbriso=47.0 .
\label{eq:supp_varyD_iso_clock}
\end{equation}
Changing the initial separation \(D\) primarily shifts the binary collision
time \(\tcollbin\).  This allows us to test whether visible-chain formation
follows the relative timing between \(\tcollbin\) and the same isolated
reference clock.

For this fixed-amplitude scan we use the same normalized timing variable
defined in Eq.~\eqref{eq:supp_eta},
\begin{equation}
\eta
=
\frac{
\tcollbin-\tbrstartiso
}{
\tbrendiso-\tbrstartiso
}.
\label{eq:supp_eta_varyD}
\end{equation}
Here \(\tbrstartiso\) and \(\tbrendiso\) are fixed to the
\(\phicen=0.0875\) isolated breathing window for every row in the separation
scan.  Thus \(\eta<0\) corresponds to a collision before the isolated
breathing window, \(0\leq\eta\leq1\) corresponds to a collision inside the
isolated window, and \(\eta>1\) corresponds to a collision after the end of
the isolated window.  As before, this isolated interval is used only as a
reference clock, not as a rigid boundary for the binary dynamics.

Table~\ref{tab:supp_separation_summary} gives the full event-time summary.  The
small-separation cases \(D=50\) and \(D=60\) collide before the isolated
breathing window and produce only weak chain candidates, with
\(S_{\rm chain,valid}^{\max}<S_{\rm vis}\).  The intermediate cases
\(D=80\) and \(D=90\) produce visible chains.  The largest separations
\(D=180\) and \(D=200\) enter the BH channel before common-remnant
formation.  They are therefore excluded by event ordering and are not
assigned a valid visible-chain score.

%---------------------------------------
\begin{table}[t]
\caption{Timing summary for the separation scan at fixed
\(\phicen=0.0875\) and \(v=0.1\). All rows are compared with the same
isolated breathing window, \([187.5,234.5]\). The normalized timing
variable \(\eta\) gives the location of the binary collision time
relative to this fixed isolated clock. The score column reports
\(S_{\rm chain,valid}^{\max}\), the maximum chain score in the valid
pre-BH classification channel. The visible-chain threshold is
\(S_{\rm vis}=2.0\). A dash indicates that the case is excluded by event
ordering because the BH diagnostic threshold is reached before
common-remnant formation.}
\begin{ruledtabular}
\begin{tabular}{cccccl}
\(D\) &
\(\eta\) &
\(t_{\rm collision}^{\rm binary}\) &
\(t_{\rm BH\,formed}^{\rm binary}
 - t_{\rm collision}^{\rm binary}\) &
\(S_{\rm chain,valid}^{\max}\) &
Classification \\
\hline
50  & \(-0.95\) & 143.0 & \(+67.0\)   & 0.999 & weak candidate \\
60  & \(-0.33\) & 172.0 & \(+75.5\)   & 0.999 & weak candidate \\
80  & \( 0.95\) & 232.0 & \(+97.5\)   & 2.898 & visible chain \\
90  & \( 1.55\) & 260.5 & \(+112.5\)  & 2.964 & visible chain \\
180 & \(10.05\) & 660.0 & \(-382.5\) & --    & BH before collision \\
200 & \(11.65\) & 735.0 & \(-482.0\) & --    & BH before collision \\
\end{tabular}
\end{ruledtabular}\label{tab:supp_separation_summary}
\end{table}
%---------------------------------------

The separation scan supports the timing-window interpretation in a way that
is complementary to the fixed-separation amplitude scan.  Since the
isolated star is fixed, the reference clock is unchanged across the table;
only the binary collision time is shifted by changing \(D\).  Collisions
that occur too early produce only weak chain candidates, while collisions
near the isolated clock produce visible chains.  For very large separations,
the binary reaches the BH diagnostic threshold before common-remnant
formation, so any later raw morphology feature is not interpreted as a
bosonic visible-chain event.

\clearpage

%==================================================================
\section{Robustness against changing the self-interaction strength}
\label{sec:supp_l500}
%==================================================================

As an additional robustness check, we repeated the fixed-separation timing
comparison for selected \(n=2\) configurations with a different quartic
self-interaction strength, \(\lambda=500\).  The binary parameters were
kept fixed at \(D=80\) and \(v=0.1\).  The same event definitions and
classification rules used in the main timing analysis were applied, and
the normalized timing variable is the same as in Eq.~\eqref{eq:supp_eta}.

\begin{figure}[t]
\centering
\includegraphics[width=0.72\textwidth]{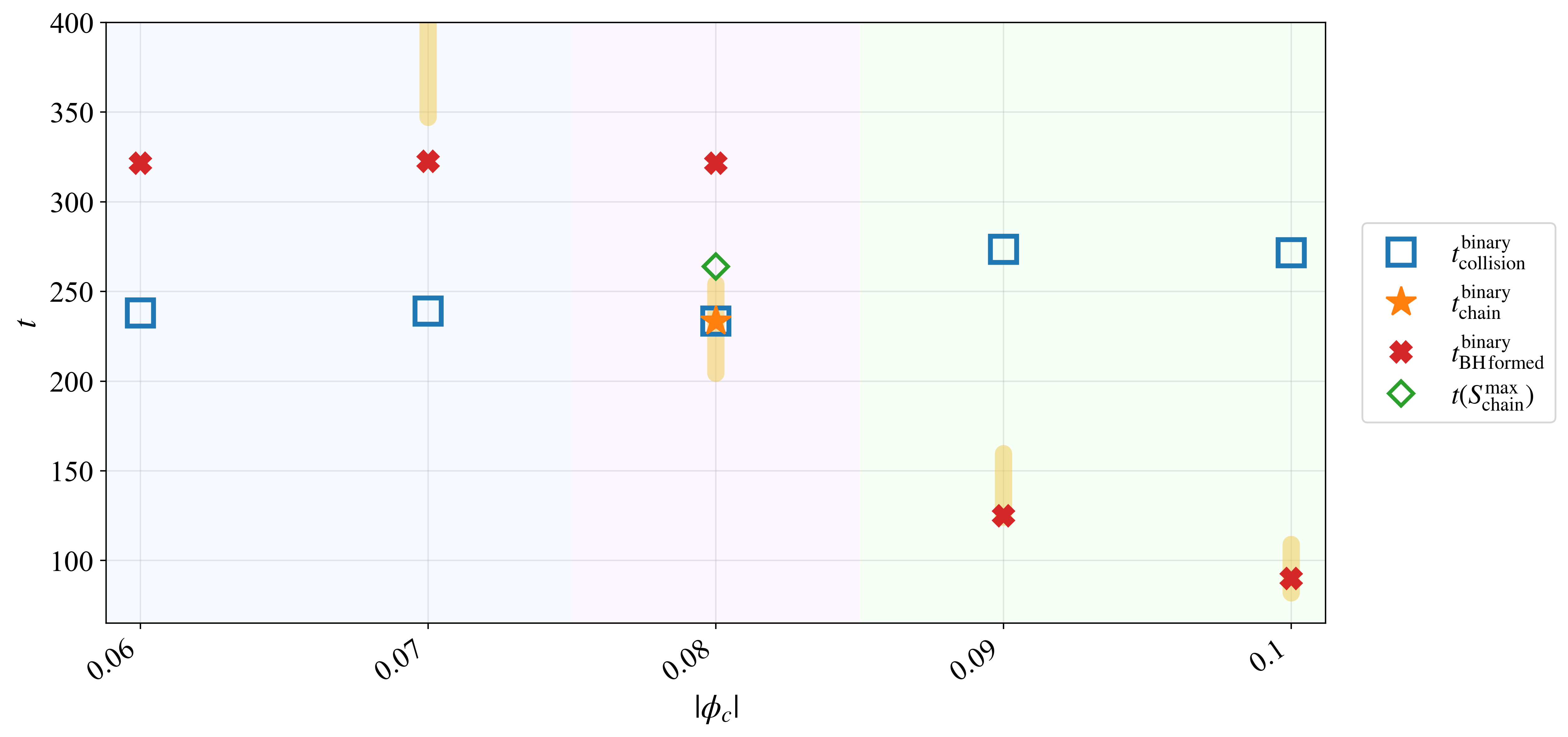}
\caption{
Additional self-interaction check for the \(n=2\), \(\lambda=500\) sequence
at fixed \(D=80\) and \(v=0.1\).  Yellow bars show the isolated breathing
intervals, blue squares mark \(\tcollbin\), orange stars mark
\(\tchainbin\), red crosses mark \(\tBHbin\), and green diamonds mark the
time at which \(S_{\rm chain,valid}\) reaches its maximum.  The same
early-collision, timing-compatible, and BH-before-collision ordering as in
the \(\lambda=400\) scan is observed.
}
\label{fig:l500_n2_binary_timescale_scan}
\end{figure}

The result is summarized in
Fig.~\ref{fig:l500_n2_binary_timescale_scan} and
Table~\ref{tab:l500_n2_timing_summary}.  The lower-amplitude cases collide
before the corresponding isolated breathing clock is reached and produce
only subthreshold chain candidates.  The intermediate case
\(\phicen=0.080\) collides inside the isolated breathing window and forms a
visible chain, with
\(\tchainbin=\tcollbin\) and
\(S^{\max}_{\rm chain,valid}>S_{\rm vis}\).  The larger-amplitude cases
reach the BH diagnostic threshold before common-remnant formation and are
therefore excluded by event ordering.

Thus changing the self-interaction strength shifts the isolated clocks and
the detailed event times, but does not change the qualitative
timing-window pattern. This supports the interpretation that the visible-chain mechanism is not an artifact of the single choice \(\lambda=400\).
%------------------------------------------------
\begin{table}[t]
\centering
\caption{
Timing-window summary for the additional \(n=2\), \(\lambda=500\)
fixed-separation scan with \(D=80\) and \(v=0.1\).  The normalized timing
variable \(\eta\) is defined in Eq.~\eqref{eq:supp_eta}.  The score column
reports \(S^{\max}_{\rm chain,valid}\), and the visible-chain threshold is
\(S_{\rm vis}=2.0\).  A dash indicates that the corresponding quantity is
not defined or that the case is excluded by event ordering.
}
\begin{ruledtabular}
\begin{tabular}{cccccccc}
\(\phicen\) &
\(\tbrstartiso\) &
\(\tbrendiso\) &
\(\eta\) &
\(\tcollbin\) &
\(\tBHbin-\tcollbin\) &
\(S^{\max}_{\rm chain,valid}\) &
Classification \\
\hline
0.060 & 596.5 & --    & --      & 238.0 & \(+83.5\)  & 0.9999 & weak candidate \\
0.070 & 347.0 & 420.5 & \(-1.47\) & 239.0 & \(+83.5\)  & 0.9997 & weak candidate \\
0.080 & 204.5 & 254.0 & \(0.59\)  & 233.5 & \(+88.0\)  & 2.943  & visible chain \\
0.090 & 124.0 & 159.5 & \(4.21\)  & 273.5 & \(-148.5\) & --     & BH before collision \\
0.100 & 82.0  & 109.0 & \(7.02\)  & 271.5 & \(-181.5\) & --     & BH before collision \\
\end{tabular}
\end{ruledtabular}
\label{tab:l500_n2_timing_summary}
\end{table}
%------------------------------------------------
%------------------------------------------------
\clearpage

%==================================================================
\section{Auxiliary external-field check}
\label{sec:supp_aux_external_field}
%==================================================================

We performed one auxiliary BS--BH evolution to check whether an
external companion field can modify the effective breathing clock of an
excited boson star.  This run is not part of the main BS--BS timing
scan and is not used in the visible-chain classification.  Its purpose
is only to provide a consistency check on the interpretation that the
binary environment can shift the detailed timing relative to the
isolated breathing window.

The boson-star component is the same \(n=2\), \(\lambda=400\)
configuration with \(|\phi_c|=0.090\) used in the fixed-separation
BS--BS comparison.  The BS--BH initial data were constructed by adapting
the initial-data prescription of Ref.~\cite{Marks:2026xvo}.  We use
Ref.~\cite{Marks:2026xvo} only for this initial-data construction; the
subsequent evolution and the breathing diagnostic are otherwise those
used throughout this work.  The black-hole mass was chosen to be
comparable to the boson-star mass, so that the companion provides a
strong external gravitational source without introducing a second scalar
configuration.

In the corresponding isolated evolution, this boson star has the
reference breathing interval
\begin{equation}
[t^{\rm iso}_{\rm br,start},t^{\rm iso}_{\rm br,end}]
=
[165.0,207.5],
\end{equation}
followed by the diagnostic BH time
\begin{equation}
t^{\rm iso}_{\rm BH}=225.5 .
\end{equation}
In the BS--BH evolution, the boson-star component enters a
breathing/tidally distorted state at \(t=172\), when the BS--BH
separation is approximately \(75.35\).  Unlike the isolated evolution,
this state does not close on the isolated-star timescale.  Instead, the
breathing/tidal distortion persists through the subsequent BS--BH
approach and into the merger stage.  No local BS-collapse diagnostic is
triggered before this stage; the minimum value of the conformal-factor
diagnostic in the monitored boson-star region remains well above the
collapse threshold, with
\begin{equation}
\chi^{\rm BS}_{\min}=0.285 .
\end{equation}

This auxiliary run is consistent with the interpretation that an
external companion field can modify the breathing/tidal-restructuring
stage of an excited boson star. It is not used as evidence for the
visible-chain classification, but only as a consistency check supporting
the use of the isolated breathing interval as a reference clock rather
than as a rigid boundary for the binary dynamics.

\clearpage
%==================================================================
\section{Constraint monitoring}
\label{sec:supp_constraints}
%==================================================================

We monitor the \(L^2\) norms of the Hamiltonian and momentum constraint
violations during both the isolated-star and binary evolutions.  This
diagnostic is not used to define the breathing window, the collision time,
or the visible-chain classification.  Its purpose is to verify that the
relevant pre-collapse and chain-formation stages occur in a controlled part
of the evolution.

\begin{figure}[t]
\centering
\begin{minipage}[t]{0.49\textwidth}
\centering
\includegraphics[width=\linewidth]{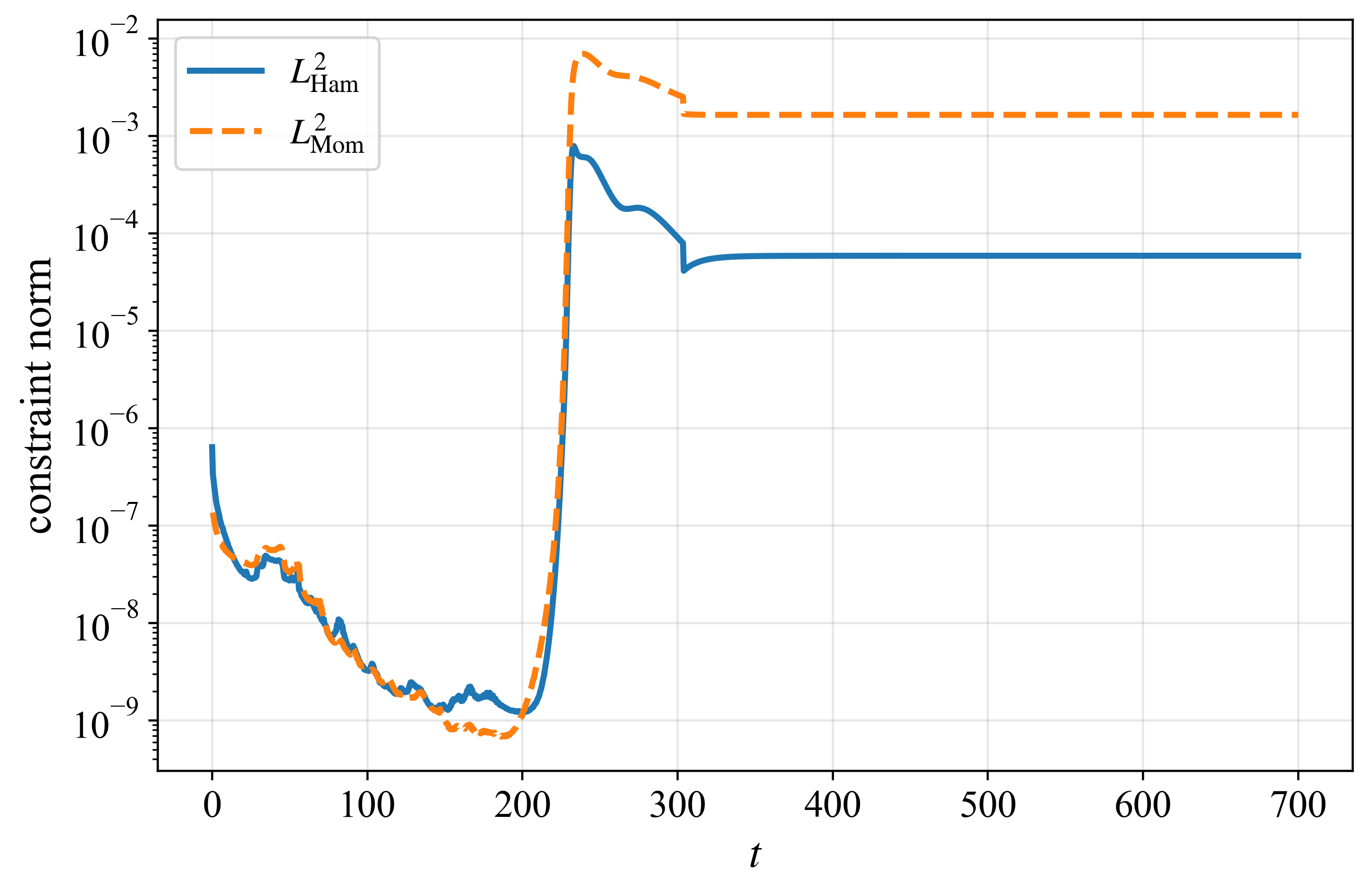}

\smallskip
\small (a) Isolated star.
\end{minipage}
\hfill
\begin{minipage}[t]{0.49\textwidth}
\centering
\includegraphics[width=\linewidth]{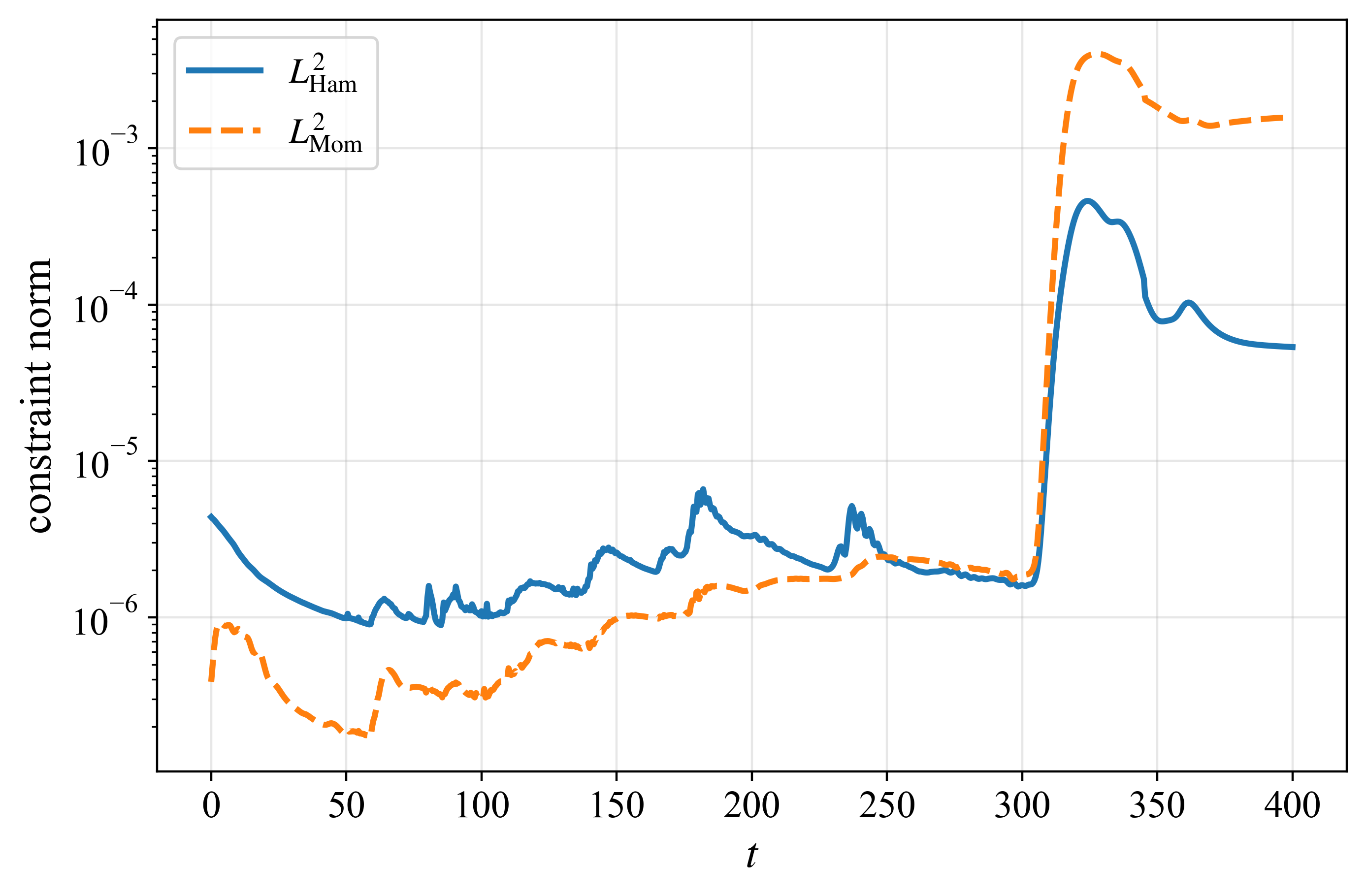}

\smallskip
\small (b) Binary run.
\end{minipage}
\caption{Constraint monitoring for the representative isolated and binary
evolutions.  Panel (a) shows the isolated \(\phicen=0.0875\) boson star.
Panel (b) shows the fiducial \(N=256\) visible-chain binary run with
\(\phicen=0.0875\), \(D=80\), and \(v=0.1\).  The early decrease of the
Hamiltonian and momentum constraint norms reflects CCZ4 damping.  The
subsequent sharp growth occurs around, or after, the BH diagnostic time; the
breathing clock and visible-chain event are identified before this
post-collapse growth.}
\label{fig:supp_constraints}
\end{figure}

Figure~\ref{fig:supp_constraints} shows representative constraint monitoring
for the isolated \(\phicen=0.0875\) star and for the fiducial binary
visible-chain run with \(\phicen=0.0875\), \(D=80\), \(v=0.1\), and
\(N=256\).  In both cases, the early decrease of the constraint norms is
consistent with the damping terms in the CCZ4 evolution system.  The sharp
growth occurs around the time at which the BH diagnostic threshold is
reached, or after that time.  The isolated breathing clock and the binary
visible-chain event are identified before this large post-collapse growth
and are therefore not inferred from the collapse-dominated part of the
evolution.

%--------------------------------------------

\clearpage
%==================================================================
\section{Resolution check}
\label{sec:supp_resolution}
%==================================================================

We checked the resolution dependence of the representative visible-chain
case \(\phicen=0.0875\), \(D=80\), and \(v=0.1\). This case is used as
the fiducial visible-chain example in the main text. The purpose of this
check is to test the robustness of the event ordering and the
visible-chain classification under refinement. As defined in the
numerical setup section, \(N\) denotes the number of grid points on the
coarsest AMR level covering the full computational domain. The fiducial
main-text resolution is \(N=256\).

Table~\ref{tab:resolution_check} shows that the resolutions
\(N\ge128\) give the same event ordering and the same visible-chain
classification. In all listed runs, within the diagnostic time resolution, the sustained
chain onset coincides with common-remnant formation,
\[
t_{\rm chain}^{\rm binary}
=
t_{\rm collision}^{\rm binary},
\]
and both occur before the diagnostic BH time. The maximum chain score
also remains above the visible-chain threshold,
\[
S_{\rm chain}^{\max} > S_{\rm vis}=2.0 .
\]
Thus the representative visible-chain classification is stable under
refinement.

We use this resolution check to establish the robustness of the event
ordering and classification, rather than to claim precision convergence
of the post-chain diagnostic BH time. The diagnostic BH time varies more
noticeably than the collision and chain-onset times, but this variation
does not affect the ordering
\[
t_{\rm chain}^{\rm binary}
\simeq
t_{\rm collision}^{\rm binary}
<
t_{\rm BH\,formed}^{\rm binary},
\]
nor does it change the visible-chain classification.

%--------------------------------------------------
\begin{table}[t]
\caption{Resolution check for the representative visible-chain run with
\(\phicen=0.0875\), \(D=80\), and \(v=0.1\). Here \(N\) is the
coarsest-level resolution defined in the numerical setup section. The
table tests the robustness of the event ordering and visible-chain
classification under refinement.}
\begin{ruledtabular}
\begin{tabular}{cccccc}
\(N\) &
\(t_{\rm collision}^{\rm binary}\) &
\(t_{\rm chain}^{\rm binary}\) &
\(t_{\rm BH\,formed}^{\rm binary}\) &
\(S_{\rm chain}^{\max}\) &
Classification \\
\hline
128 & 235.00  & 235.00  & 335.00  & 2.766 & visible chain \\
192 & 232.667 & 232.667 & 332.667 & 2.895 & visible chain \\
256 & 232.00  & 232.00  & 329.50  & 2.898 & visible chain \\
512 & 231.75  & 231.75  & 321.50  & 2.899 & visible chain \\
\end{tabular}
\end{ruledtabular}\label{tab:resolution_check}
\end{table}

\clearpage
%==================================================================
\section{Three-dimensional morphology check}
\label{sec:supp_3d_check}
%==================================================================

\begin{figure}[t]
\centering
\includegraphics[width=0.72\textwidth]{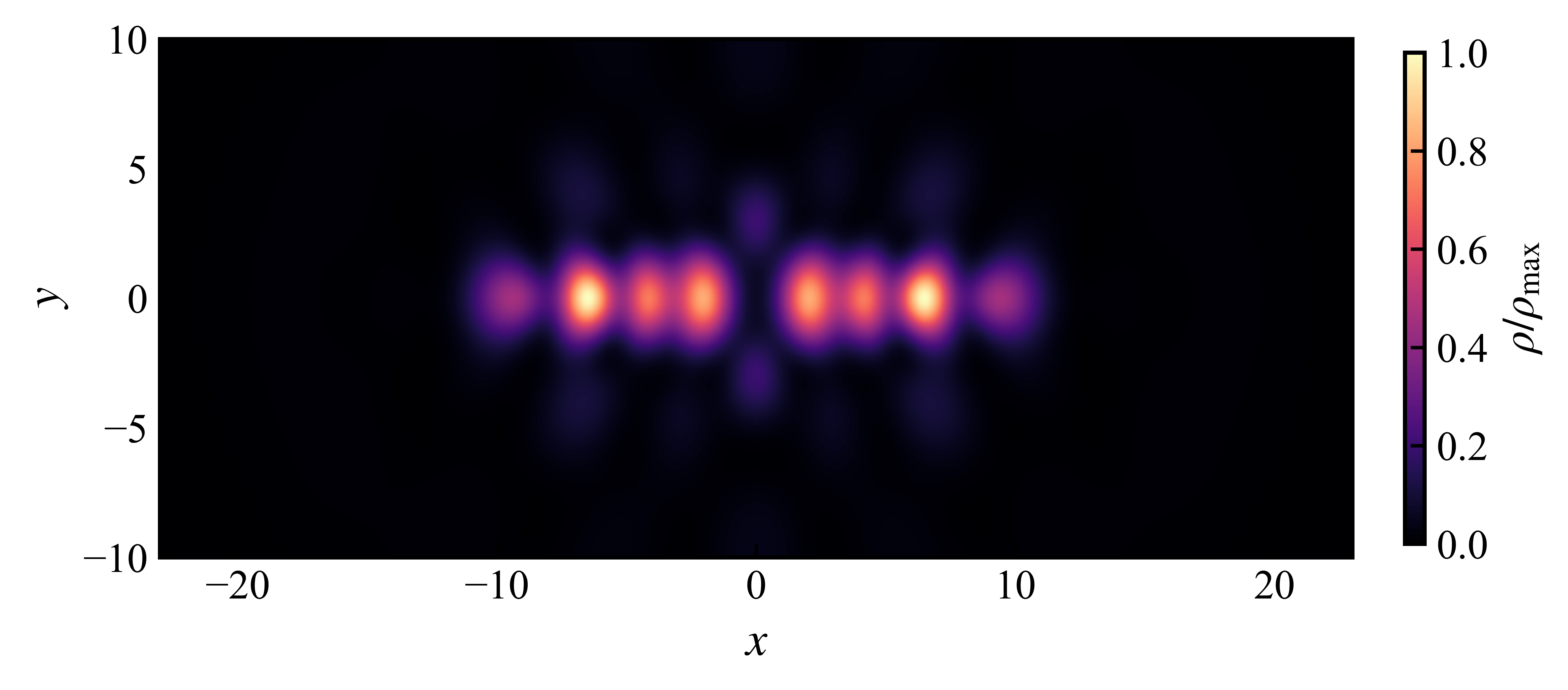}
\caption{Density slice from the 3D evolution of the representative
visible-chain case with \(\phicen=0.0875\), \(D=80\), and \(v=0.1\), shown
at \(t=275\).  The colour scale shows the density normalized by the maximum
value in the displayed slice.  The slice exhibits an axis-supported
multi-peak structure consistent with the morphology seen in the
symmetry-reduced evolution.}
\label{fig:supp_3d_snapshot}
\end{figure}

As a visual cross-check of the symmetry-reduced morphology, we also evolved
the representative visible-chain case with the three-dimensional code.
Figure~\ref{fig:supp_3d_snapshot} shows a density slice at \(t=275\) for the
same initial data, \(\phicen=0.0875\), \(D=80\), and \(v=0.1\).  The 3D
evolution exhibits an axis-supported multi-peak remnant that is consistent
with the morphology seen in the modified-Cartoon evolution.  This comparison
is not used as an independent classification criterion.  It is included only
as a qualitative check that the morphology illustrated in the main text is
not an apparent feature introduced by the two-dimensional half-plane
visualization.

%----------------------------------------------------

%==================================================================
\bibliography{BS}